\documentclass[12pt,preprint]{aastex}

\def\xmm{{XMM-{\it Newton}}}
\def\chandra{{\it Chandra}}
\newcommand{\cgs}{ ${\rm erg~cm}^{-2}~{\rm s}^{-1}$} 
\newcommand{\lum}{\rm erg~s$^{-1}$}
\def\farcs{\hbox{$.\!\!^{\prime\prime}$}}
\slugcomment{Submitted to ApJ}

\shorttitle{}
\shortauthors{Civano et al.}

\begin{document}

\title{A Runaway Black Hole in COSMOS: Gravitational
Wave or Slingshot Recoil?}

\author{F. Civano\altaffilmark{1}, M. Elvis\altaffilmark{1},
G. Lanzuisi\altaffilmark{1,2}, K. Jahnke\altaffilmark{3}, G.
Zamorani\altaffilmark{4}, L. Blecha\altaffilmark{5},
A. Bongiorno\altaffilmark{6}, M. Brusa\altaffilmark{6}, 
A. Comastri\altaffilmark{4}, H. Hao\altaffilmark{1},
A. Leauthaud\altaffilmark{7}, A. Loeb\altaffilmark{5},
M. Mignoli\altaffilmark{4}, V. Mainieri\altaffilmark{8},
E. Piconcelli\altaffilmark{9}, M. Salvato\altaffilmark{10},
N. Scoville\altaffilmark{10}, J. Trump\altaffilmark{11},  
C. Vignali\altaffilmark{12}, 
T. Aldcroft\altaffilmark{1}, M. Bolzonella\altaffilmark{4},
E. Bressert\altaffilmark{1},
A. Finoguenov\altaffilmark{6,21}, A. Fruscione\altaffilmark{1},
A. M. Koekemoer\altaffilmark{13},
N. Cappelluti\altaffilmark{6}, F. Fiore\altaffilmark{9},
S. Giodini\altaffilmark{6}, R. Gilli\altaffilmark{4},
C. D. Impey\altaffilmark{11}, S.~J. Lilly\altaffilmark{14},
E. Lusso\altaffilmark{4,12}, S. Puccetti\altaffilmark{15},
J. D. Silverman\altaffilmark{14}, H.~Aussel\altaffilmark{16},
P. Capak\altaffilmark{10}, D. Frayer\altaffilmark{10}, E. Le
Floc\'h\altaffilmark{17}, H.~J.~McCracken\altaffilmark{18},
D. B. Sanders\altaffilmark{17}, D. Schiminovich\altaffilmark{19},
Y. Taniguchi\altaffilmark{20} }

\altaffiltext{1}{Harvard Smithsonian Center for astrophysics, 60 Garden St., Cambridge, MA 02138}
\altaffiltext{2}{Dipartimento di Fisica, Universit\`a di Roma La Sapienza, P.le A. Moro 2, 00185 Roma, Italy }
\altaffiltext{3}{Max Planck Institut f\"ur Astronomie, K\"onigstuhl 17, Heidelberg, D-69117, Germany}
\altaffiltext{4}{INAF-Osservatorio Astronomico di Bologna, via Ranzani 1, I-40127 Bologna, Italy}
\altaffiltext{5}{Astronomy Department, Harvard University, 60 Garden St., Cambridge, MA 02138}
\altaffiltext{6}{Max Planck Institut f\"ur extraterrestrische Physik Giessenbachstrasse 1, D--85748 Garching, Germany}
\altaffiltext{7}{LBNL \& BCCP, University of California, Berkeley, CA 94720, USA}
\altaffiltext{8}{European Southern Observatory, Karl-Schwarzschild-Str. 2, D-85748 Garching, Germany}
\altaffiltext{9}{INAF-Osservatorio Astronomico di Roma, via di Frascati 33, I-00040 Monteporzio Catone, Italy} 
\altaffiltext{10}{California Institute of Technology, MC 105-24, 1200 East California Boulevard, Pasadena, CA 91125}
\altaffiltext{11}{Steward Observatory, University of Arizona, 933 North Cherry Avenue, Tucson, AZ 85721, USA}
\altaffiltext{12}{Dipartimento di Astronomia, Universit\`a degli Studi di Bologna, Via Ranzani 1, I-40127 Bologna, Italy}
\altaffiltext{13}{Space Telescope Science Institute, 3700 San Martin Drive, Baltimore, MD 21218, USA}
\altaffiltext{14}{Institute of Astronomy, Swiss Federal Institute of Technology (ETH H\"onggerberg), CH-8093, Z\"urich, Switzerland}
\altaffiltext{15}{ASI Science Data Center, via Galileo Galilei, 00044 Frascati Italy}
\altaffiltext{16}{ AIM Unit\'e Mixte de Recherche CEA CNRS, Universit\'e Paris VII UMR n158, Paris, France} 
\altaffiltext{17}{Institute for Astronomy, University of Hawaii, 2680 Woodlawn Drive, Honolulu, HI, 96822}
\altaffiltext{18}{Institut d'Astrophysique de Paris, UMR 7095 CNRS, Universit\'e Pierre et Marie Curie, 98 bis Boulevard Arago, F-75014 Paris, France} 
\altaffiltext{19}{Department of Astronomy, Columbia University, MC2457,550 W. 120 St. New York, NY 10027}
\altaffiltext{20}{Research Center for Space and Cosmic Evolution, Ehime University, Bunkyo-cho, Matsuyama 790-8577, Japan}
\altaffiltext{21}{University of Maryland, Baltimore County, 1000 Hilltop Circle, Baltimore, MD 21250, USA}

\begin{abstract}

We present a detailed study of a peculiar source detected 
in the COSMOS survey at z=0.359. Source CXOC~J100043.1+020637, also
known as CID-42, presents two compact optical sources embedded in the same
galaxy. The distance between the two, measured in
the HST/ACS image, is 0.495"$\pm$0.005" that, at the redshift of the source,
corresponds to a projected separation of 2.46$\pm$0.02 kpc. A large ($\sim$1200
km/s) velocity offset between the narrow and broad components of H$\beta$ has
been measured in three different optical spectra from the VLT/VIMOS and
Magellan/IMACS instruments. CID-42 is also the only X-ray source in COSMOS having in
its X-ray spectra a strong {\it redshifted} broad
absorption iron line, and an iron emission line, drawing an {\it inverted}
P-Cygni profile. The \chandra\ and \xmm\ data show that the absorption line is 
variable in energy by $\Delta$E=500 eV over 4 years and that the
absorber has to be highly ionized, in order not to leave a signature in the soft
X-ray spectrum. That these features, the morphology, the velocity offset and the {\it inverted}
P-Cygni profile, occur in the same source is unlikely
to be a coincidence. We envisage two possible explanations, both exceptional,
for this system: (1) a gravitational wave recoiling black hole (BH), 
caught 1-10 Myr after merging, (2) a Type 1/ Type 2 system in the same galaxy where the Type~1
is recoiling due to slingshot effect produced by a triple BH system. The
first possibility gives us a candidate gravitational
waves recoiling BH with both spectroscopic and imaging signatures. In the second
case, the X-ray absorption line can be explained as a BAL-like outflow
from the foreground nucleus (a Type 2 AGN) at the rearer one (a Type 1 AGN),
which illuminates the otherwise undetectable wind, giving us the first
opportunity to show that fast winds are present in obscured AGN, and possibly 
universal in AGNs. 
\end{abstract}


\section{Introduction}

Double super-massive black holes (SMBHs) within a single galaxy are predicted
from the combination of hierarchical models of structure formation (see Colpi \&
Dotti 2009 for a recent review), and the observed link between black 
hole (BH) mass and host galaxy
bulge mass in the local Universe (e.g., Magorrian et al. 1998, Kormendy \&
Gebhardt 2001). Kiloparsec scale double nuclei should be common at redshift
z$\sim$2 where the merging process is very efficient (Springel et al. 2005) but
at this high z they are hard to resolve. At later epochs (z$<$0.7) they are
easier to find, but the merger rate is lower (Hopkins et al. 2008) and binary
SMBHs should be relatively scarce. For these reasons, observational evidence
for binary active SMBHs at kpc separations remains sparse. A few low redshift examples have been
found. The most famous double active SMBHs are the highly obscured pair in the
Ultra-luminous Infrared galaxy (ULIRG) NGC 6240 (Komossa et al. 2003;
z=0.0245). \chandra\ imaging and spectroscopy of NGC 6240 show the presence of
two active galactic nuclei (AGNs) separated by just 1.2 kpc. 
Only few other AGN pairs at kiloparsecs separations have been
discovered so far, all by chance: Arp 299 (Ballo et al. 2004; z=0.01), Mrk 463
(Bianchi et al. 2008; z=0.05), 3C 75 (Hudson et al. 2006; z=0.02), 
and 0402+379 (Rodriguez et al. 2006; z=0.05). A candidate
sub-parsec binary system at z=0.38 has been recently discovered in the SDSS
through the detection of two emission line systems in the optical spectrum
(Boroson \& Lauer 2009), but this interpretation is still under debate (e.g., Chornock
et al. 2010).  

Once the SMBH binary tightens, its merger is characterized by the emission of
gravitational waves (GW) and the merged BH under
favorable conditions (of galaxy mass ratio and BH spins) can recoil with 
respect to the center of the galaxy (Peres 1962, Bekenstein 1973) at substantial 
velocities (few$\times$10$^2$-10$^3$ km/s, Campanelli et al. 2007a). 
If not enough mass is driven to the SMBH binary before the merger, it can stall
until the arrival of a third galaxy induces the hardening of the binary which 
ejects the just arrived BH for gravitational slingshot effect (Saslaw et al. 1974). 

Recently, observational searches were initiated for
single quasars which are displaced from their host galaxy spatially or
spectroscopically due to GW recoil after the merger (e.g., Bonning et al. 2007).
Only a few candidates have been reported so far (Komossa et al. 2008,
Shields et al. 2009), all of them discovered through the detection of multiple
line systems in optical spectra, but more prosaic interpretations for these
candidates are not yet ruled out.
The quasar HE 0450-2958 is candidate to be a BH ejected by the nucleus of a
companion galaxy during a major merger (Magain et al. 2005, Hoffman \& Loeb
2006), but questions remain about the nature of this source (e.g., Merritt et al. 2006, 
Kim et al. 2007, Jahnke et al. 2009a).

A survey for binary SMBH systems or the discovery of many recoiling BH would
be helpful to test evolutionary merger models and to put more robust constraints
on our understanding of BH and galaxy co-evolution (e.g., Hopkins et al. 2008).

The Cosmic Evolution Survey (COSMOS; Scoville et al. 2007a) is a deep
multiwavelength survey of 2 deg$^2$, based on a 590-orbit ACS program (Koekemoer
et al. 2007), the largest mosaic ever made by the HST. Of the $10^6$ galaxies to
$i_{AB}\sim 27$ detected in the full field, the $i$-band (F814W) ACS data
provide morphological information down to $i_{AB} < 23$ for 70,000 galaxies and
enables detailed structural analysis out to $z \sim 0.8$ for $\sim$16,000 galaxies. The
large number of spectra ($>$20,000, Lilly et al. 2007, 2009, Trump et al. 2007) 
available for galaxies and quasars allow to search for sources with more than 
one line system, although the typical resolution available ($R=600-700$) only 
allows sources with high velocity shifts to be found. The rich multiwavelength 
COSMOS database allows detailed
study of the source SEDs at all frequencies from radio to X-ray. These properties
make COSMOS an excellent database in which to look for sources just before or
after the SMBHs binary merging. According to the prediction of Volonteri \&
Madau (2008), the number of detectable off-center AGN (up to 30 kpc away from the
center of its host) in the HST COSMOS area should be 
$\sim$30, for the best case of large kicks (spinning holes), long decay
timescales (no bulge), and long active phase, while considering those X-ray
emitting the expected number is $\sim$1. In contrast, the short lifetime of
binary BH at small separations implies a much lower number of
sub-parsec candidates in the whole area, even at low redshift
($<$1/deg$^2$, Volonteri et al. 2009).

SMBHs are easier to find when they are active, and X-ray surveys are the most
efficient means for finding AGNs (e.g., Brandt \& Hasinger 2005). Hence, we started
our survey for pre- and post-binary candidates with the optical counterparts of
X-ray sources within COSMOS.

A visual inspection of the optical counterparts of all 2600 X-ray sources,
detected in the \chandra-COSMOS (C-COSMOS, Elvis et al. 2009) and XMM-COSMOS (Hasinger et
al. 2007) surveys, using the HST/ACS F814W images (Koekemoer et al. 2007;
Leauthaud et al. 2007), led to the discovery of just one candidate recoil SMBH in 
the source CXOC~J100043.1+020637 (CID-42, Elvis
2009) at z=0.359. At the same time, Comerford et al. (2009) reported the discovery of dual AGNs in the same
source, through the detection of two emission line systems in a Keck/DEIMOS
spectrum.
 
We report here a detailed study of CID-42, including the SED analysis and optical and 
X-ray imaging and spectroscopy. We discuss several models for the nature of this system.

In this paper, magnitudes are reported in the AB system (Oke \& Gunn 1983) and a
WMAP 5-year cosmology\footnote{
 http://lambda.gsfc.nasa.gov/product/map/dr3/params/lcdm\_sz\_lens\_wmap5.cfm}
with H$_0$ =71~km~s$^{-1}$~Mpc$^{-1}$, $\Omega_M$ = 0.26 and
$\Omega_{\Lambda}$=0.74 is assumed.

\section{The X-ray Source CXOC~J100043.1+020637}

\begin{figure*}
\includegraphics[width=\textwidth]{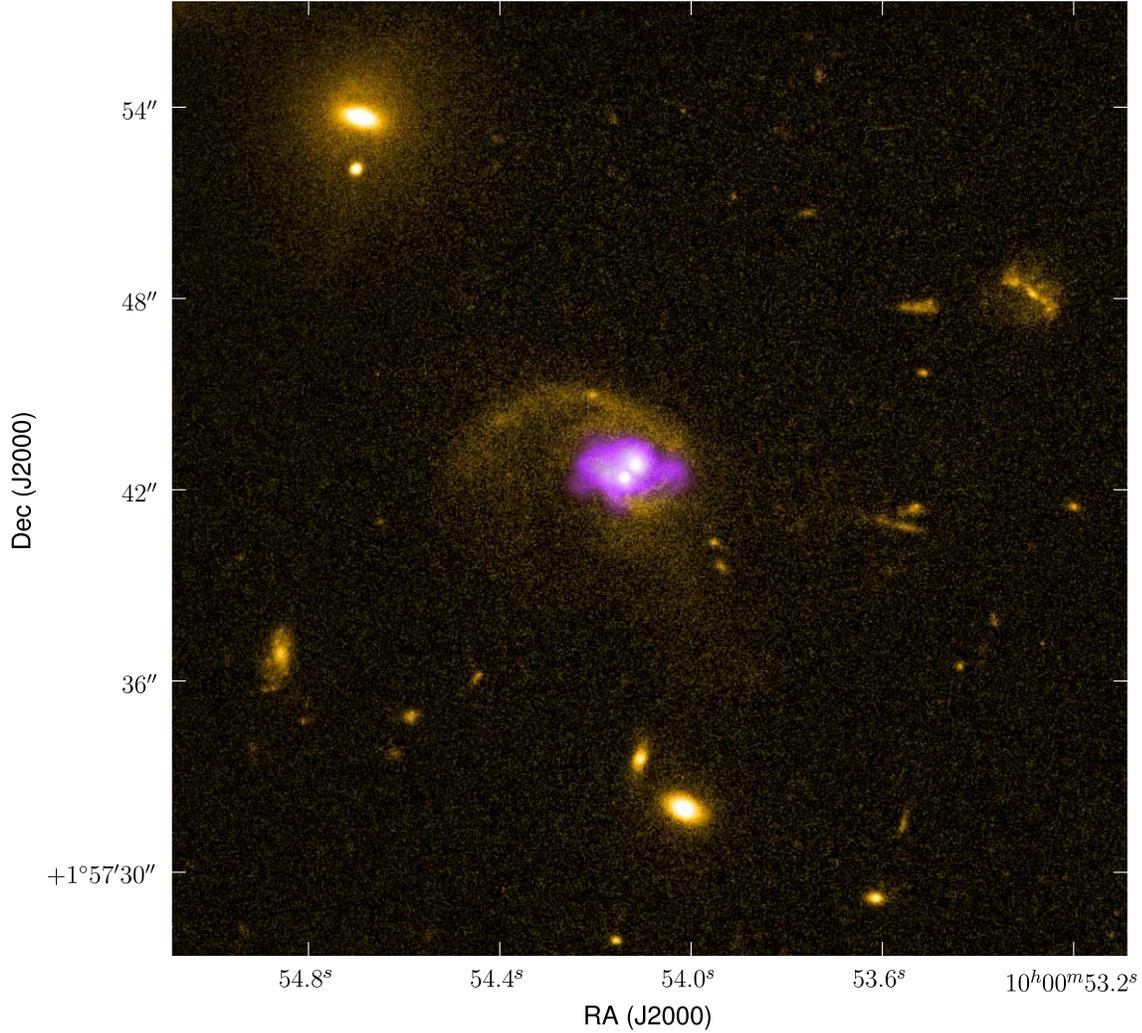}
\caption{\small Two band 30''$\times$30'' image of the double nucleus
source. North is up, East is to the left. The image has been made summing the \chandra\
ACIS 0.5-8 keV (purple color scale) and ACS F814W (yellow/white color scale)
images. The ACS data was stretched with a logarithmic function. The high
resolution of the ACS image (0.03''/pixel) allows to resolve the two sources. The
X-ray data was optimized using the GREYCStoration algorithm
(http://cimg.sourceforge.net/greycstoration/). }\label{3col}
\end{figure*}

\begin{figure*}[t]
\includegraphics[width=0.5\textwidth]{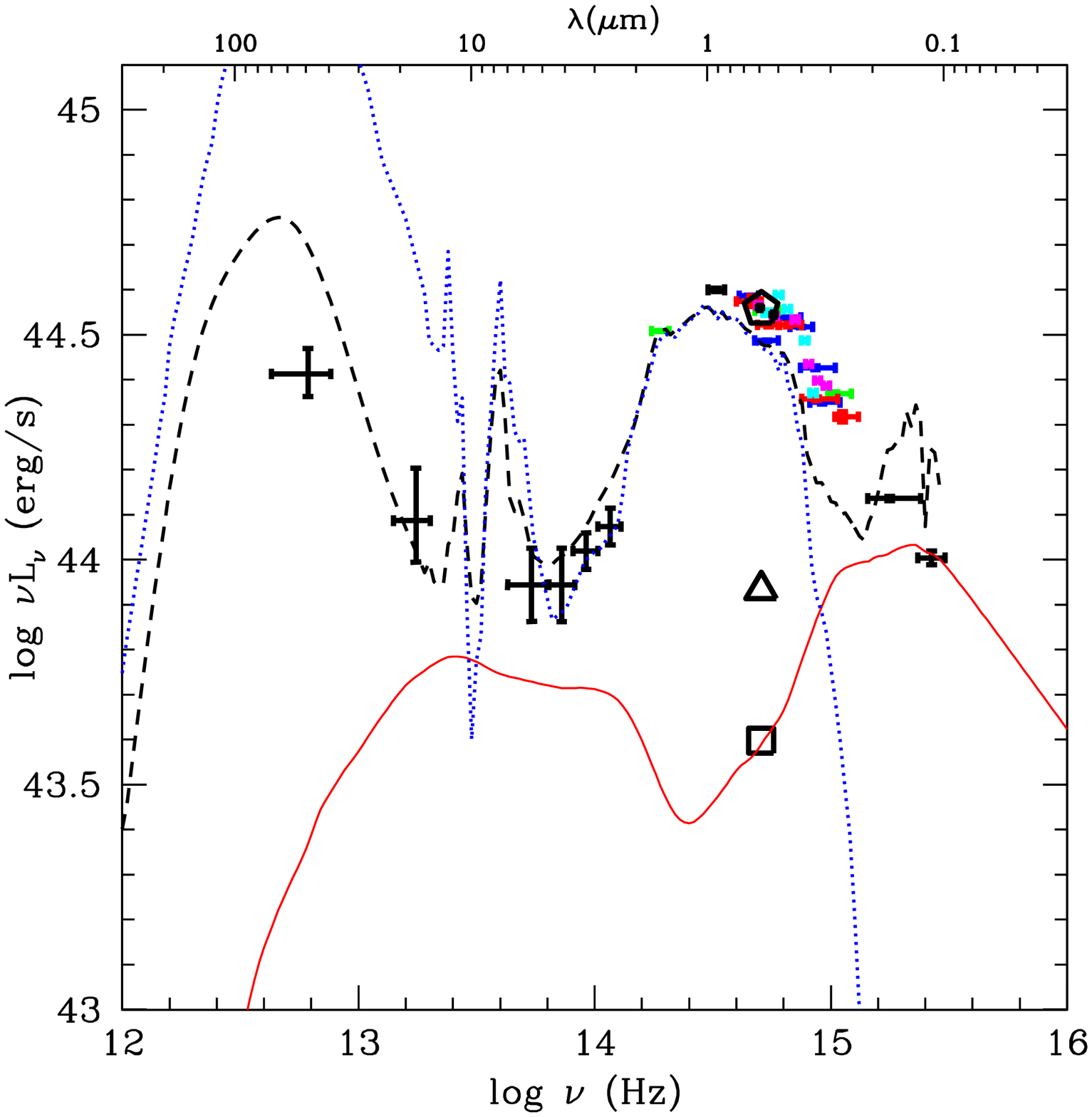}
\includegraphics[width=0.5\textwidth]{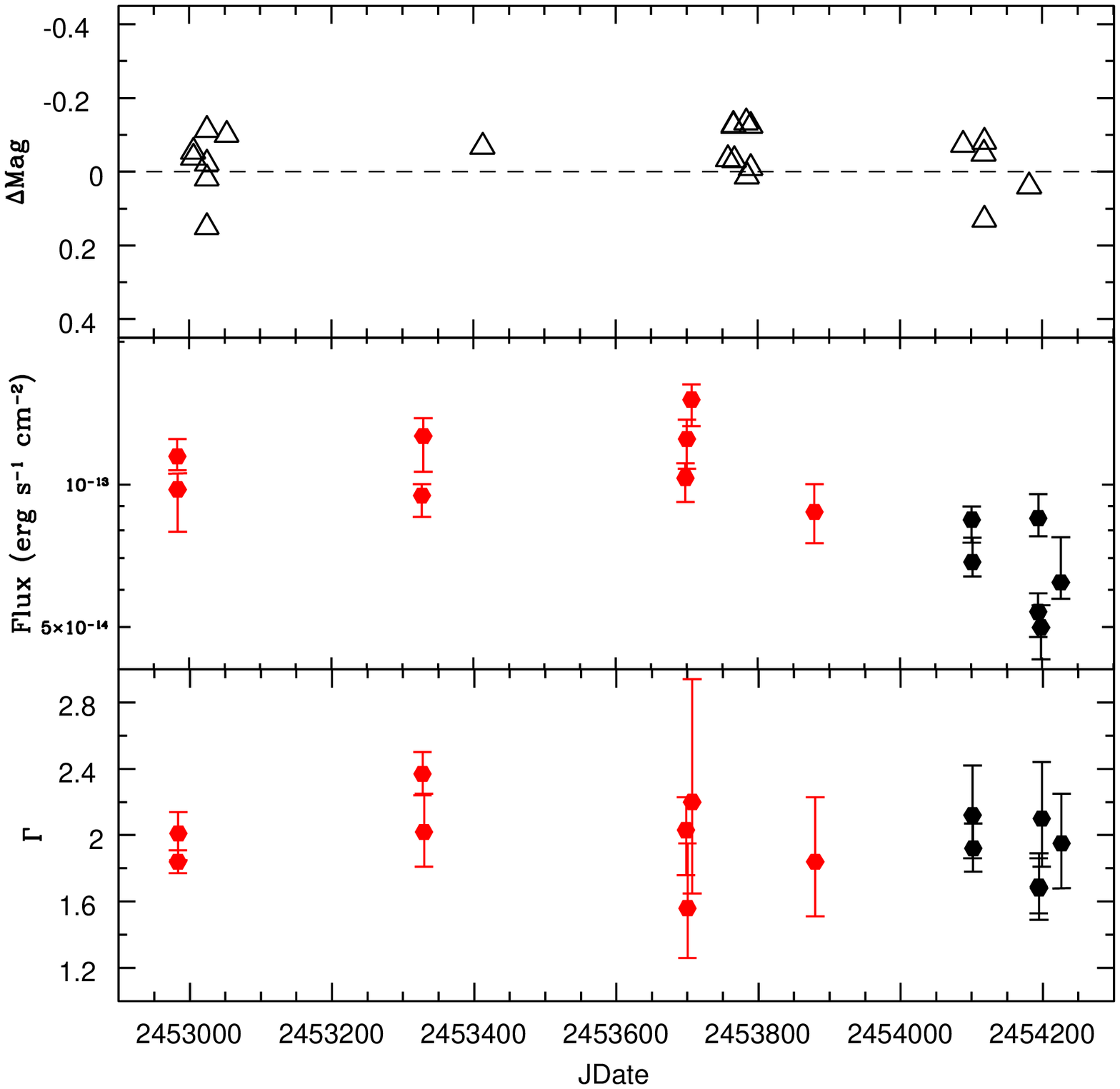}
\caption{\small {\it Left}: Infrared to UV rest-frame SED. The small points
represent all the COSMOS data for source CID-42 with 1 $\sigma$ errors: MIPS 70 and 24$\mu m$,
IRAC, CFHT K, J and i, Subaru broad and narrow bands, SDSS, CFHT u and
Galex. The big symbols represent the luminosity computed for the two compact sources
separately (square = SE source, triangle = NW source) and for the total system
(pentagon symbol) using the surface brightness decomposition technique (Section
3). The Elvis et al. (1994) radio quiet quasar SED (continuous red line) has
been normalized to the ACS luminosity ($\lambda$5900 \AA\ in the rest frame) 
of the SE source. The dashed and dotted
lines are a Seyfert 1.8 and NGC 6240, respectively (Polletta et al. 2007). {\it
Right}: {\it Top}: Time variability of the optical flux (as computed by Salvato
et al. 2009). In the optical, CID-42 was observed during 4 epochs (Subaru
broad-band, CFHT i band, the first set of intermediate Subaru bands, and the
second epoch of intermediate Subaru bands). {\it Center}: X-ray full band light
curve (red=\xmm, black=\chandra). {\it Bottom}: X-ray spectral slope versus 
time. }
\label{sed}
\end{figure*}

The peculiar source is the optical counterpart of the X-ray source CID-42 
(Elvis et al. 2009; Civano et al. 2010 in prep.), also known as XMMU~J100043.1+020636 
(Cappelluti et al. 2009; Brusa et al. 2007, 2010 submitted). Figure~\ref{3col} shows a 30''$\times$30'' two color image, created by
overlaying the \chandra\ 0.5-8 keV (in purple color scale) and ACS F814W (in
yellow/white color scale) images,
around CID-42. Two sources (white) are clearly visible in the HST
image. The galaxy shows structures to the NE and SW suggestive of tidal tails formed during a
first galaxy-galaxy encounter (e.g., Cox et al. 2006). 

CID-42 is in the central area of the COSMOS field and, for this
reason, has been observed in all the COSMOS bands, multiple times in some cases, allowing
the construction of the FIR-UV spectral energy distribution (SED). In
Fig.~\ref{sed} (left panel), the total rest-frame SED (using data from Frayer et
al. 2009, Le Floch et al. 2009, Sanders et al. 2007, Capak et al. 2007,
McCracken et al. 2010, Taniguchi et al. 2007, Zamojski et al. 2007) is compared
with the template of a Seyfert 1.8 and of the double nucleus prototype NGC 6240
(dashed and dotted line, respectively; Polletta et al. 2007) normalized to the
K-band luminosity. The Elvis et al. (1994) quasar SED, normalized as it will be
explained in Section 3.1,
is also shown (continuous red line). The radio emission is 0.138$\pm$0.038 mJy at 20
cm (Schinnerer et al. 2007, Bondi et al. 2008), and the source is classified as
radio-quiet using the Wilkes \& Elvis (1987) definition ($R_L = log(f_{5GHz}/f_{B})$=-0.55). 
Although the optical spectrum shows
strong emission lines typical of an AGN (see Section~\ref{spec}), the optical
SED is dominated by the galaxy component. The infrared
luminosity is lower than that of NGC 6240 (normalized in the K band) but
consistent with that of a Luminous Infrared Galaxy (L$_{24\mu m}
\sim$3.2$\times$10$^{10}$~L$_{\odot}$) which could suggest ongoing
star-formation activity as expected in a merger phase (Mihos \& Hernquist
1996). However, nuclear activity could be contributing strongly to the IR emission,
as shown from the comparison with the Seyfert 1.8 SED. The good fit to the UV
with the last SED suggests that the obscuring gas is not Compton Thick as in NGC~6240. The
bolometric luminosity estimated from the SED is L$_{24\mu
m-UV}$=2.86$\times$10$^{11}$~L$_{\odot}$. 
The stellar mass and star formation rate (SFR) of the host galaxy,
derived from the fit to the {\it u} band to 4.5$\mu m$ photometry (to avoid the wavelengths where the
AGN contribution can be relevant) using a set of 11 models with
smooth star formation histories (Bruzual and Charlot 2003; IMF from Chabrier
2003), are M$_{star} \sim 2.5\times10^{10}$ M$_{\odot}$ and SFR$\sim$100 
M$_{\odot}$/yr (Bolzonella et al. 2009). Our estimated M$_{star}$ and
SFR are meant to be upper limits because the nuclear component can still
contribute to the above quantities. Given this mass and SFR, CID-42 is
actively star-forming, with a SFR consistent with the overall X-ray selected AGN
population (Silverman et al. 2009).

The 5 year optical and near-IR light curve, as computed by Salvato et al.
(2009), is shown in Fig.~\ref{sed} (top-right). No significant variability has been
detected, as expected given that the optical SED is dominated by the host galaxy
light.

The source was imaged in the X-rays for 74 ks by \xmm\ and for 165 ks by
\chandra\ ACIS-I. The \xmm\ data were taken over 2.5 years (8 observations)
and the \chandra\ data over 1.5 years (6 observations), for a total of 14
observations over 4 years. The reported exposure is the sum of all the effective
exposure times. The luminosities derived in the two co-added sets of
observations are L$_{0.5-10keV}$=4.8$\times$10$^{43}$ \lum\ and
1.9$\times$10$^{43}$ \lum\ for \xmm\ and \chandra, respectively. The factor 2.5
difference between the two luminosities is too large to be explained by
cross-calibration problems between the two instruments and implies
significant X-ray variability. The full 4 year X-ray light curve
(Fig.~\ref{sed}, right-middle) shows a factor of 3 variability in the full
(0.5-10 keV) band flux, while in the soft (0.5-2 keV) and hard (2-10 keV) band
the variability reaches a factor 4 and 4.5, respectively.
No significant variability in the spectral slope has been measured (Fig.~\ref{sed}, right-bottom).

\begin{figure*}[ht]
\centering{ \includegraphics[width=0.3\textwidth]{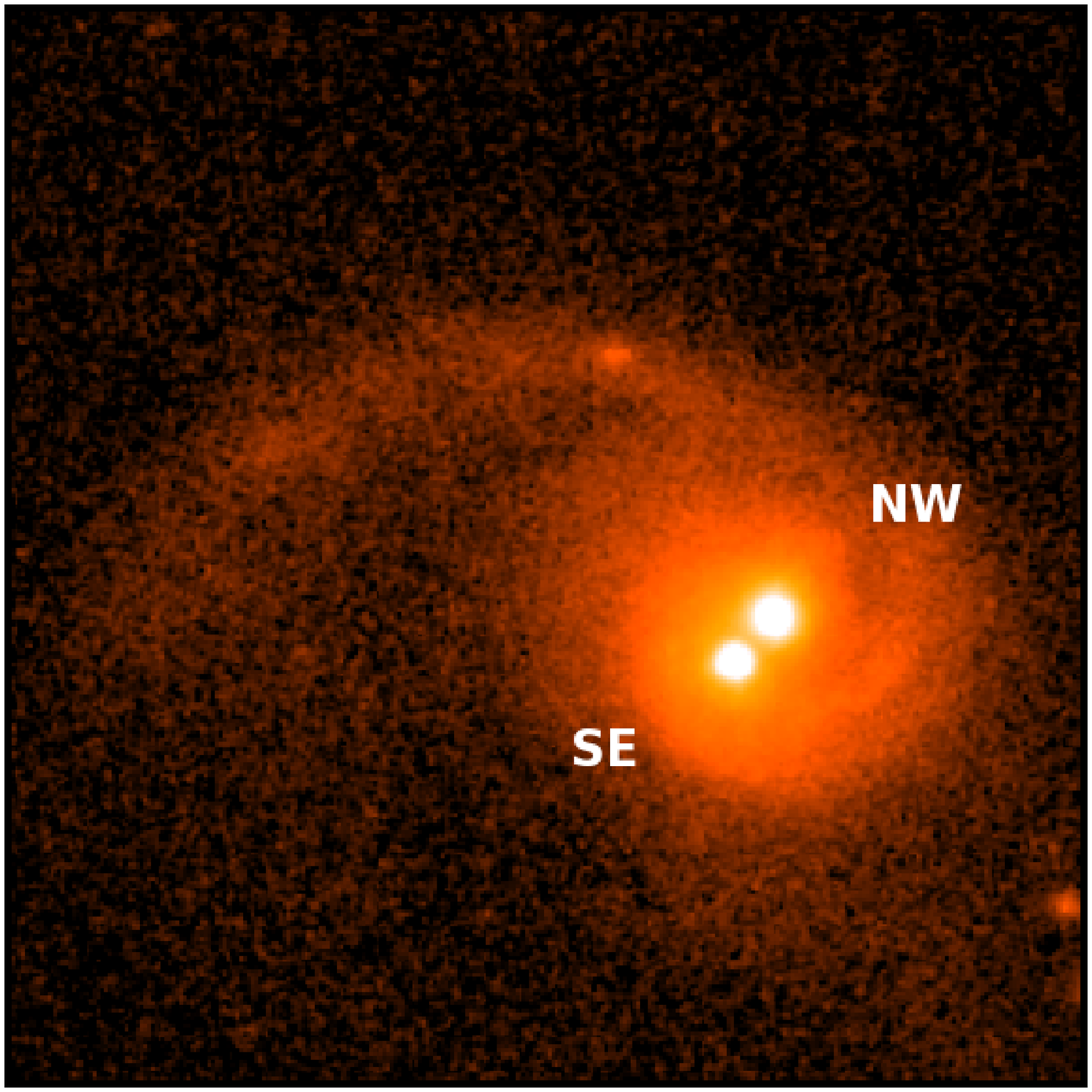}
\includegraphics[width=0.3\textwidth]{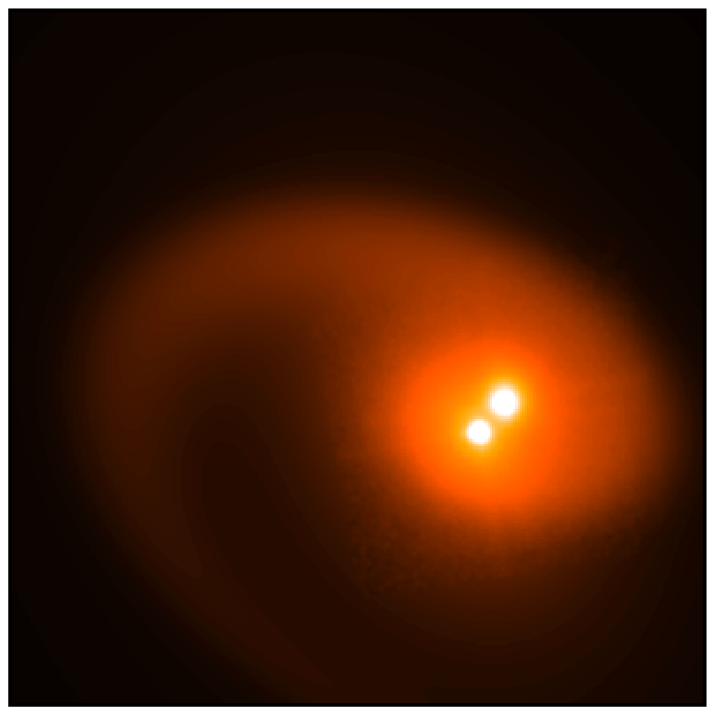}
\includegraphics[width=0.3\textwidth]{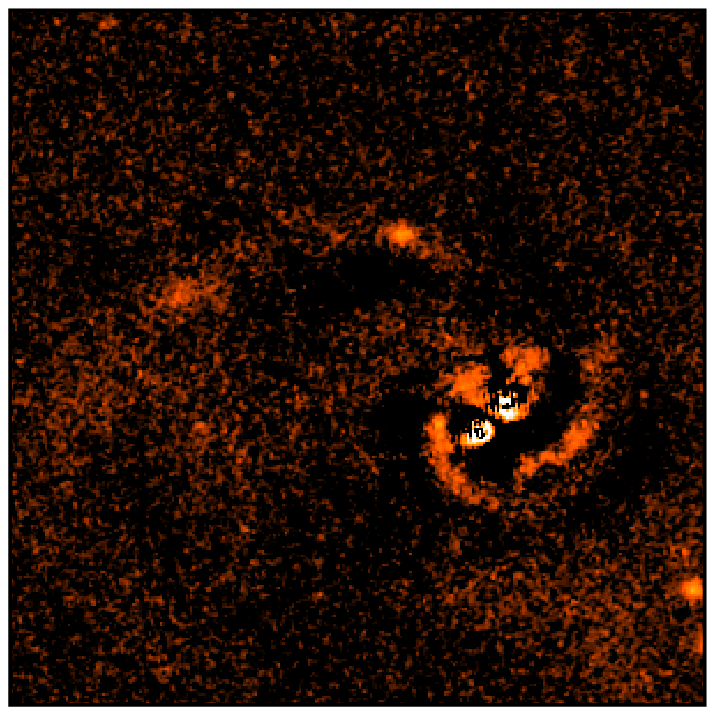}}
\caption{\small Surface brightness decomposition results: input 9''$\times$9'' ACS image (left), model used 
(1 point like, 3 Sersic components; middle) and residuals 
obtained from the fitting (right). The images are shown in a logarithmic scale. North is up, East is 
to the left. }
\label{hst_fit}
\end{figure*}

\section{Imaging analysis}
\subsection{HST Imaging analysis}
\label{hst}
Figure~\ref{hst_fit} (left) shows a 9''$\times$9'' zoom around source CID-42 in
the ACS/HST image (0.03''/pixel; filter F814W). The distance measured between
the two optical centers is 0.495"$\pm$0.005" (16.5 pixels) which corresponds to
a projected separation of 2.46$\pm$0.02~kpc at the source's redshift.

We modeled the optical surface brightness of CID-42 with GALFIT (Peng et
al. 2002, Version 3.0\footnote{C. Y. Peng private communication.}). We use an
empirical point-spread-function (PSF) created by averaging stars in the ACS
frames (Jahnke et al. 2004, 2009b) to represent a point source in the field and
to use for convolution of the parametric galaxy models.

We tested several combinations and numbers of point sources and Sersic profiles
(Sersic 1968) with a free Sersic-parameter $n$. We also made use of two new
features in GALFIT V3.0, a Fourier mode modification of isophotes, restricted in
this case to F1 (lopsidedness), and a mode to model the NE tidal arm. The goals
of this exercise were to (1) test whether the two nuclei in the system are point
sources, (2) extract reliable point source fluxes, and (3) put constraints on
the relative fluxes of the galaxy components around the two sources.

The best description of the system consists of one point source and three Sersic
components. The model and the residuals are shown in Figure~\ref{hst_fit}
(middle and right). We find that, above the general background, only the
south-eastern nucleus (SE) is well modeled with a point source, while the
north-western nucleus (NW) is extended though compact. This finding is
independent of the exact number and geometry of large scale galaxy components added to the
model. By this model, we estimate the point-like SE nucleus to have
$F814W=20.51$~mag (square in Fig.~\ref{sed}, left) with an uncertainty of
$\sim0.1$~mag. The first Sersic component corresponds to the NW nucleus and is
very compact, with a scale length of $\sim$0\farcs1 ($\sim$ 0.5 kpc), smaller
than the typical dimension of host galaxies of X-ray selected AGN (e.g., Schade
et al. 2000), and $F814W=19.67$~mag (triangle in Fig.~\ref{sed}, left). The second
Sersic component is a large spheroid (r$_{eff}$=3'', $\sim$15 kpc) and the third
Sersic component describes part of the overall light distribution plus the
spiral arm. The latter
two components combined have $F814W=18.6$~mag and together describe the host
galaxy light distribution, likely extending the profile of the NW source. The
overall system brightness obtained with the fit is $F814W=18.11$~mag (pentagon
in Fig.~\ref{sed}, left), in agreement with the photometry used for the SED.

In order to test the goodness of the fit, we ran the GALFIT analysis with a
simple model (1 point-like source and 1 Sersic component), as was reported by
Gabor et al. (2009). This fitting runs into calculation boundaries, indicating
that a 2 component model is not adequate to represent this complex system. A
minimum of 3 components (1 PSF and 2 Sersic) is required. Instead, the best fit
is obtained using 4 components, as explained above.

Given the point-like unresolved profile of the SE source, it could be considered
as a unobscured BH (i.e. Type 1 AGN) and it is possible to compute its bolometric luminosity, by
normalizing to its F814W magnitude a typical Type 1 AGN SED (Elvis et al. 1994;
red line in Figure \ref{sed}, left). This gives a bolometric luminosity for the SE
source of L$_{24\mu m-UV}$=2.8$\times$10$^{44}$ \lum, consistent with being a
moderately luminous AGN.

\subsection{\xmm\ Imaging Analysis}

The merging system CID-42 is not an isolated source, as is evident in
Fig.~\ref{3col}. Using combined \xmm\ and \chandra\ data (with prior point
source removal), extended emission has been detected around CID-42. The X-ray
detected group (Finoguenov et al. 2007, 2010 in prep.) is spectroscopically
confirmed with 7 galaxy members in the zCOSMOS bright catalog (Lilly et
al. 2007, 2009) and 30 members in the photometric redshift catalog (Ilbert et
al. 2009), of which 10 are visible in Fig.~\ref{3col}. The extended
X-ray luminosity in the 0.1-2.4 keV band is 1.4$\times$10$^{42}$ \lum. The
X-ray centroid, though uncertain (30'' positional error), is consistent with the
position of CID-42. The detection of extended X-ray emission from hot gas of
the group allows us to estimate the mass of the halo in which the merging galaxy
resides. The total gravitational mass of the group is M$_{200}$= 2.5$\pm$0.7
$\times$10$^{13}$ M$_{\odot}$\footnote{M$_{200}$ is the mass enclosed in
R$_{200}$, which is the radius within which the matter density is 200 times the
critical density of the Universe.}, based on the average Lx-M$_{200}$
calibration of Leauthaud et al. (2010). The corresponding R$_{200}$ is 106''
($\sim$0.5 Mpc), so CID-42 is located in the core of the group. The source
CID-42 has been identified also as the second most massive group galaxy (MMGG,
details in Leauthaud et al. 2010). Furthermore, the galaxy group hosting CID-42
is embedded in a large over-dense region at z$\sim$0.37 covering the whole COSMOS
field (Scoville et al. 2007b, Gilli et al. 2009).

\subsection{Chandra Imaging Analysis}

CID-42 was imaged with \chandra\ in 6 ACIS-I pointings with different exposure
times (14 to 46 ks) and off-axis angles (3'-7.5') spread over 4 months. Given
the \chandra\ ACIS-I resolution (0.5"/pixel), it is not possible to perform an
accurate decomposition analysis as performed with HST/ACS data, but it is
possible to study the emission in different energy bands.

We analyzed the X-ray emission in both soft (0.5-2 keV) and hard (2-7 keV) band
images, using the observation in which the source has been observed at the
smallest off-axis angle ($\sim$ 3') and with the longest exposure time (46 ks).
In this observation, the source has 187 net counts in the full band in a 
region corresponding to 95\% of the encircled energy fraction for a point source 
at that position. 

\begin{figure}[t]
\centering\includegraphics[width=0.45\textwidth]{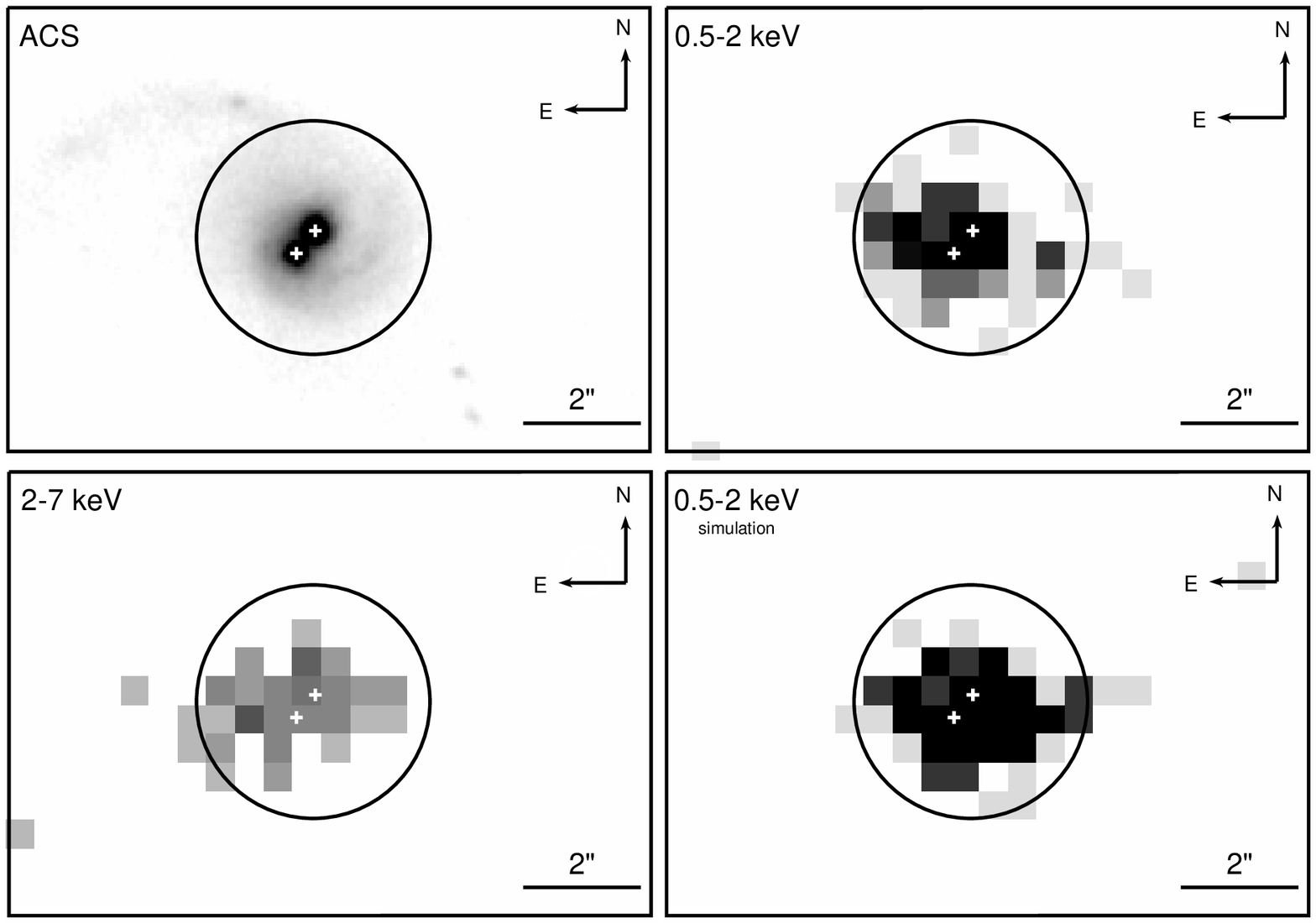}
\caption{\small {\it From top left to bottom-right}: HST/ACS image; ACIS-I images 
in the 0.5-2 keV, 2-7 keV energy bins and simulated image in the 0.5-2 keV 
band assuming point-like emission from the two nuclei. 
The 4 images are matched on the same coordinate scale. The crosses represent 
the position of the optical nuclei, and the circles show a 2'' radius to help in 
comparing the images. }
\label{ds9}
\end{figure}

Most of the counts (123 out of 187) are detected in the 0.5-2 keV band.
Comparing to the PSF which applies for this energy band, off-axis angle, and
azimuth, there is some evidence for extension along the axis joining the two
optical nuclei. To characterize this better we did a detailed ray-trace
simulation using ChaRT\footnote{http://cxc.harvard.edu/chart/} and
MARX\footnote{http://space.mit.edu/CXC/MARX/} with the assumption of equal flux
from the two optical nuclei. The results are illustrated in Fig.~\ref{ds9}
where we show the \chandra\ soft band X-ray image (top right) and the simulation
(lower right). The observed and simulated images are qualitatively consistent,
indicating that there is no evidence for galaxy-scale X-ray diffuse emission in
this source. One would like to usefully constrain the flux ratio of the two
point sources, but this is not possible here given the PSF extent and
astrometric uncertainty at 3' off-axis. For completeness, we also simulated the
2-7 keV hard band data as coming from the two nuclear point sources and also in
this case we found that the observed image (Fig.~\ref{ds9}, lower left) is
consistent with the simulated image.

\begin{figure}[h]
\centering\includegraphics[width=0.45\textwidth]{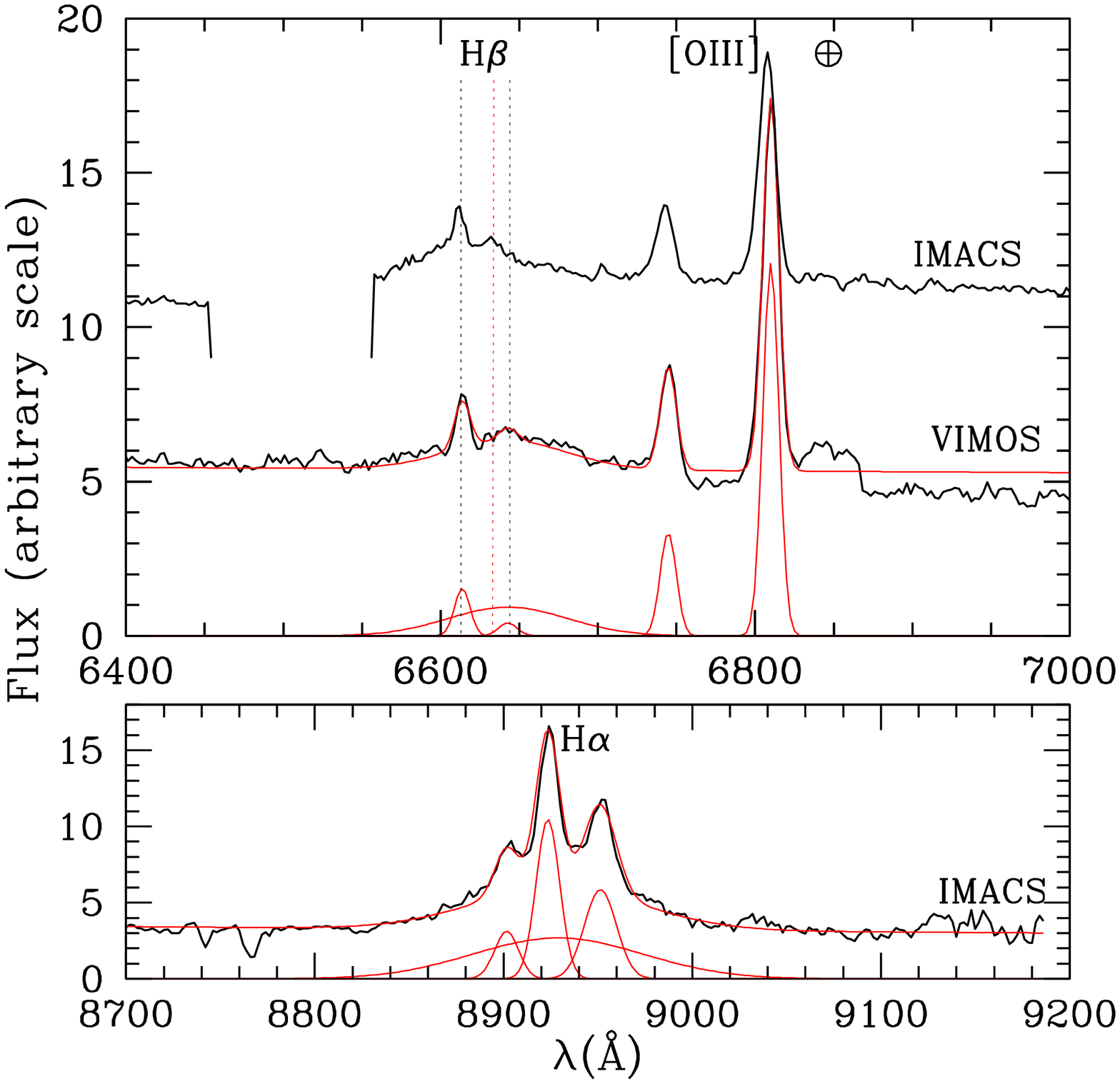}
\caption{\small {\it Top}: CID-42 zCOSMOS and IMACS spectra (black line) around 
the H$\beta$ line plotted in the observed frame. The components used to reproduce 
the spectral features in the zCOSMOS spectrum 
are shown at the bottom of the frame in red, as well as the spectral fit. The flux 
is in arbitrary unit. The emission lines used in the fit are labelled. The chip gap in the IMACS spectrum 
is evident. The thin dashed lines represent the observed wavelength of the zCOSMOS 
H$\beta$ lines (black) and the predicted wavelength where the broad H$\beta$ should peak in the IMACS
spectrum (red). The strong residuals at 6825 \AA\ are due to the presence of an absorption telluric line. 
{\it Bottom}: CID-42 IMACS spectrum around the H$\alpha$ line plotted in the observed 
frame. The line colors are as above. }
\label{optspec}
\end{figure}

\section{Spectroscopic Analysis}

\subsection{Optical spectroscopy} 
\label{spec}
Optical spectra are available from SDSS (Adelman-McCarthy et al. 2008),
IMACS/Magellan (Trump et al. 2007, 2009a) and VIMOS/VLT (zCOSMOS, Lilly et
al. 2007, 2009). The source has been observed on February 2005 with IMACS and
then twice with VIMOS on May and June 2005. Given the 1'' width of the slits
(oriented N-S) both in the VIMOS and IMACS observations, and the small E-W
separation of the sources ($\sim$0.3''), the flux from both falls within 
the slit. The 2D spectra have been analyzed, but none of them spatially resolve
the two nuclei, while in Comerford et al. (2009) the two nuclei seem to be
resolved in a 2D DEIMOS spectrum; the spatial resolution of the DEIMOS spectrum 
is similar to the IMACS one (IMACS: 1 pixel=0.11'', DEIMOS 1 pixel=0.11'', VIMOS 1 pixel=0.205'') but
its spectral resolution (R$\sim$3000) is higher. The two zCOSMOS spectra (5500-9500
\AA) have an average signal-to-noise ratio of 6 (11 around the H$\beta$ line;
Fig.~\ref{optspec}) and spectral resolution of R$\sim$600. The IMACS spectrum
(5600-9200 \AA, R$\sim$700) has a detector chip gap at 6450-6560 \AA, cutting
off part of the H$\beta$ line. The H$\alpha$ emission line is in a region where
fringing becomes strong in VIMOS. Although H$\alpha$ is visible in the IMACS
spectrum, which uses the nod and shuffle technique to avoid fringing, the peak
of the line is saturated.

For the above reasons, we started the spectral analysis by fitting the blue part
of the zCOSMOS spectrum around the H$\beta$ and the [OIII] doublet. The fitting
was performed using the IRAF routine SPECFIT (Kriss 1994). We modeled the
spectrum with a linear component for the continuum plus 4 Gaussian lines, 
1 broad and 3 narrow: the
[OIII] doublets, the broad and narrow H$\beta$. The
central wavelength, width and intensity of the gaussians were left 
free to vary. The redshift
measured from the peak of the narrow emission lines ([OIII] and H$\beta$) is 
z=0.359, coincident with the redshift listed in Lilly et al. (2007, 2009) and
Comerford et al. (2009). This redshift is also consistent with the one measured
from absorption lines (e.g., H$\gamma$). The FWHM of the narrow emission lines, corrected for
instrumental broadening, is 560$\pm$120 km/s.

\begin{figure*}
\centering
\fbox{\includegraphics[width=0.9\textwidth]{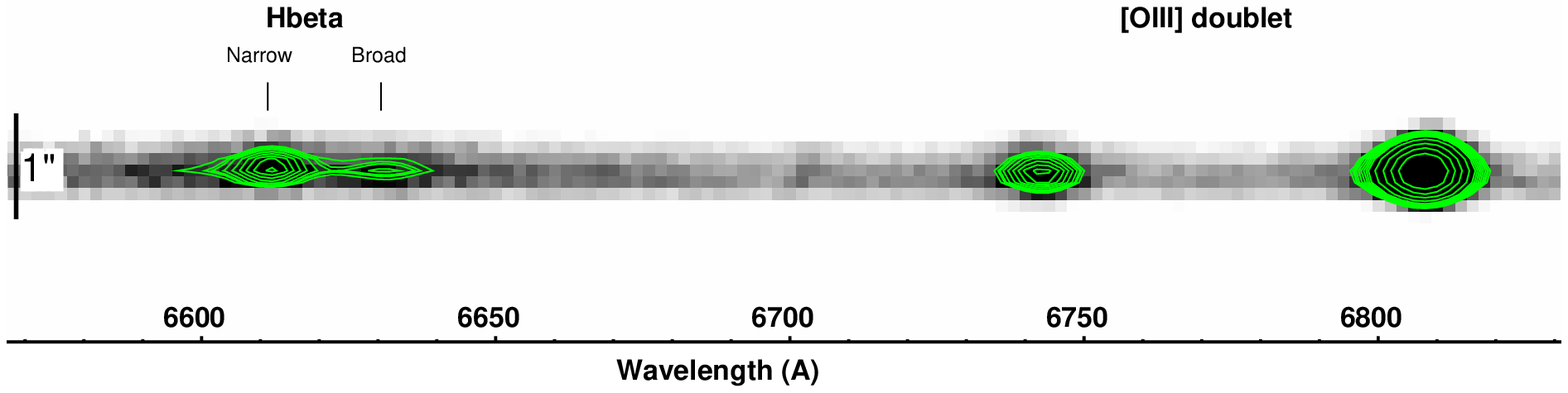}}
\caption{Two-dimensional IMACS spectrum of source CID-42: zoom around H$\beta$ 
and [OIII] doublet region. The contours have been produced using SAOImage DS9 analysis tools. }
\label{bidim}
\end{figure*} 

The 3 optical spectra show a typical Type 1.5 (Osterbrock \& Ferland 2006) AGN feature:
a broad H$\beta$ component plus a narrow core that has a line width similar to
the forbidden [OIII] lines. Unusually though, the broad and narrow H$\beta$ components
of CID-42 have widely offset peaks, potentially implying that the emission comes
from different regions, consistent with the presence of two unresolved sources.
The same offset is measured in both the zCOSMOS spectra and so, in order to have
a better measurement of this offset, we summed them. We measure an offset
$\Delta \lambda \sim$ 30$\pm$7~\AA\ (1360$\pm$320 km/s) between the centroids of
the broad and narrow H$\beta$. Small residuals at the peak of the broad line
possibly suggest the presence of another narrow component at the same
wavelength that, when fitted, unfortunately has a flux below the S/N ratio of
the spectrum.

If the offset between the two components is due to the narrow lines being
produced by one of the two optical sources and the broad line by the other, then
an instrumental shift due to the spatial separation and the orientation of the
sources with respect to the E-W VIMOS dispersion axis has to be considered.
This effect shifts the broad component towards the red by 4~\AA, which should be
subtracted from the measured offset. The net offset (26$\pm$7 \AA) implies a
difference in velocity of $\sim$1180$\pm$320 km/s between the H$\beta$ narrow
and broad components, under this assumption.

The 2D IMACS spectrum (see Fig. \ref{bidim}) shows a clear shift of the peak of
the broad line with respect to the narrow line, but does not show the spatial
separation as seen in Comerford et al. (2009), despite the same pixel scale of
the two instruments (IMACS and DEIMOS), most likely due to different seeing
conditions.
 
The presence of a detector chip gap close to the H$\beta$
region in the IMACS spectrum prevents a good fitting of the line profile.
However, a narrow line is present with a broad peak redshifted from the
narrow core. Using the shift measured in the zCOSMOS spectrum, we predicted the
wavelength where the broad line in IMACS should peak. The instrumental shift due
to position and slit orientation in the IMACS spectrum would move the broad
component towards the blue (i.e. in the opposite direction to the zCOSMOS
spectrum shift) because of the opposite dispersion direction of the two instruments, so that
the two peaks have a different observed offset with respect to zCOSMOS. The
predicted offset in IMACS is $\sim$20 \AA, which matches with the observed
line peak (red dashed line in Fig.\ref{optspec}, top). If the lines are not
coming from two different sources, the predicted offset in IMACS would not
correspond with the observed line peak, although it could be consistent within
the errors.

We performed further checks to test the possible presence of a second system of
lines with the same offset of the broad H$\beta$. The first would be the
detection of a second [OIII] doublet with the same shift measured in the
H$\beta$ but, at the expected wavelength, a Telluric feature ($\sim$6825 \AA)
prevents the detection of the corresponding redshifted [OIII] lines. The second
test would be the detection of a redshifted broad H$\alpha$ line. As explained
above, H$\alpha$ lies in a region dominated by fringing in the zCOSMOS
spectrum, while in IMACS the line peak is saturated. The fit of the H$\alpha$
and [NII] doublet in the IMACS spectrum returns 3 narrow lines at the same
redshift of the narrow [OIII] doublet (Fig.\ref{optspec}, bottom). The broad
H$\alpha$ component peak is slightly offset ($\sim$ 5 \AA) to the red from the
narrow one but this shift is not enough to be consistent with the shift
predicted from H$\beta$, even when the instrumental shift is taken into account.
Although a proper fitting of the continuum has been performed, a 1:2 ratio for
the [NII] doublet flux is observed instead of the 1:3 ratio required by atomic
physics. On the contrary, the expected 1:3 ratio has been measured in the [OIII]
doublet flux in all spectra. This suggests that, around H$\alpha$, the IMACS spectrum
is not well behaved and so the measurement of the broad line peak is still uncertain.

We also performed a simple two-component fit to the IMACS spectrum to determine
the relative nuclear and host contributions in the 4000-9000 \AA\ range.
Following Trump et al. (2009b), we find the combination of quasar (Vanden Berk
et al. 2001) and galaxy (Eisenstein et al. 2001) which maximizes the
Bayesian probability function and best describes the spectrum. The estimated
nuclear to host ratio is $\sim$3:2. Systematic errors, arising from neglecting 
the contribution from other components, could affect this estimated ratio. 
However, this simple fit shows that the spectrum is not
consistent with a simple blue quasar, but instead includes a significant host
contribution (consistent with the GALFIT decomposition of Section 3.1). The
significant host contribution may also be part of the reason for which the 
fitting around the H$\alpha$ region returns different results from those
obtained for H$\beta$ and [OIII].

\begin{figure*}[!ht]
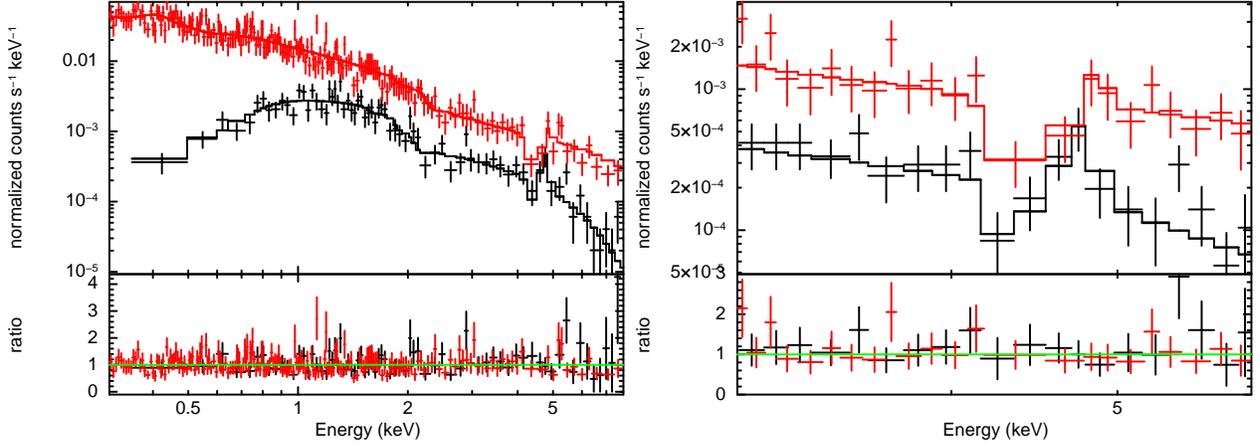

\includegraphics[ angle=-90,width=0.5\textwidth]{figure6l.ps}
\includegraphics[ angle=-90,width=0.5\textwidth]{figure6r.ps}
\caption{\small {\it Left}: X-ray spectra of CID-42 (\xmm=red, \chandra=black)
fitted with an absorbed power-law plus 2 Gaussian lines. The spectra have been
rebinned for presentation purposes at 5 counts per bin. The spectra are folded
with the instrumental response. {\it Right}: Zoom of
the X-ray spectra at the energies around the absorption and emission features
fitted with a broad and a narrow Gaussian line, respectively. The {\it
inverted} P-Cygni profile is visible. }
\label{xspec}
\end{figure*}

\subsection{X-ray Spectral Analysis}

The X-ray data available ($\sim$ 800 counts in \chandra\ ACIS and $\sim$ 2200
counts in the \xmm\ pn) allow us to perform good quality X-ray spectral analysis
when \xmm\ pn and \chandra\ ACIS spectra are fitted together. The results are
reported in Table \ref{Xray}. Due to the modest number of counts, we used the Cash
statistic (Cash 1979). A fit with an absorbed power law returns a spectral
index $\Gamma$=2.11$\pm$0.06, with no cold intrinsic absorption, in good
agreement with the results of Mainieri et al. (2007, $\Gamma$=2.18$\pm$0.07).
Residuals in the very soft energy band ($<$0.5 keV, mostly in the \xmm\
spectrum) suggest the presence of a thermal component ($k$T$<$0.13 keV) that
accounts for 8\%-9\% of the 0.3-2 keV luminosity, which could be due to either a
nuclear soft-excess or hot gas in the galaxy. The addition of the soft
component to the fit produces a slightly flatter spectral slope with
$\Gamma$=1.95$_{-0.06}^{+0.07}$. 

The most striking features of the X-ray spectrum are around 4.5~keV (observed
frame, Fig.~\ref{xspec}, left). The residuals suggest the presence of both
emission and absorption features in both spectra, forming a kind of {\it
inverted} P-Cygni profile, i.e., the absorption component is {\it redshifted}
with respect to the emission component (Fig.~\ref{xspec}, right).

The emission feature, which is consistent with being a neutral iron line, has a
constant flux and is more prominent in the \chandra\ spectrum (EW=570$\pm$260
eV, in the rest frame), where the continuum is fainter, than in the \xmm\ one
(EW=142$_{-86}^{+143}$ eV, in the rest frame), where the continuum is 2 times higher.

The absorption feature instead shows strong variability between \xmm\ and
\chandra\ data, both in flux and width: the line is well fitted with a broad
($\sigma$=0.22$^{+0.33}_{-0.11}$ keV, 1$\sigma$ error) Gaussian line in the
\xmm\ spectrum, where it is more prominent, while it is unresolved in the
\chandra\ data ($\sigma <$0.5 keV). Assuming z=0.359, the central rest-frame
energy of the absorption line is around 6 keV and therefore it seems to be a {\it
redshifted} absorption iron line.

The shape and intensity of the background spectrum have been studied in order to
discard the possibility for the absorption line to be a background artifact. 
The fact that the absorption line has been detected in spectra from 2
different instruments (EPIC and ACIS), already rejects this hypothesis.

In order to check if the {\it inverted} P-Cygni feature can be mimicked by other
spectral models, we also performed the spectral fitting using more complex
models, a broken power-law and a partial covering model. The two models reproduce the
continuum better than a single power-law, given the larger number of parameters,
but none of them reproduces the absorption and emission features. In particular,
the edge around 5-6 keV in the partial covering model does not reproduce the
absorption feature. The $\Delta$Cash statistic obtained adding the two
Gaussian lines to the models is the same as that obtained with a single
power-law, meaning that the absorption feature is real and independent from the model
adopted to reproduce the continuum. We note that the partial covering model returns 
N$_H\sim$10$^{23}$ cm$^{-2}$ and a covering fraction of 40\%, which can be
translated as the sum of two power-law of almost the same intensity, one
obscured and the other not.

A spectral fitting with a photo-ionized gas model to the total \xmm\ spectrum
has also been performed. Unfortunately, given the low CCD resolution, 
it is not possible to distinguish single absorption lines in the soft
band. The equivalent width of the absorption feature with a photoionized-gas
model implies a high column density of the absorber (N$_H\sim$5$\times$
10$^{23}$cm$^{-2}$), and the modest absorption measured in the soft band (N$_H
<$0.9$\times$ 10$^{20}$cm$^{-2}$ in the rest frame) requires the absorber to be highly ionized
(log~$\xi\sim$3). From the centroid of the line, we measure a velocity of the 
outflow of 30000$\pm$4000 km/s.

\begin{figure}[t]
\includegraphics[width=0.5\textwidth]{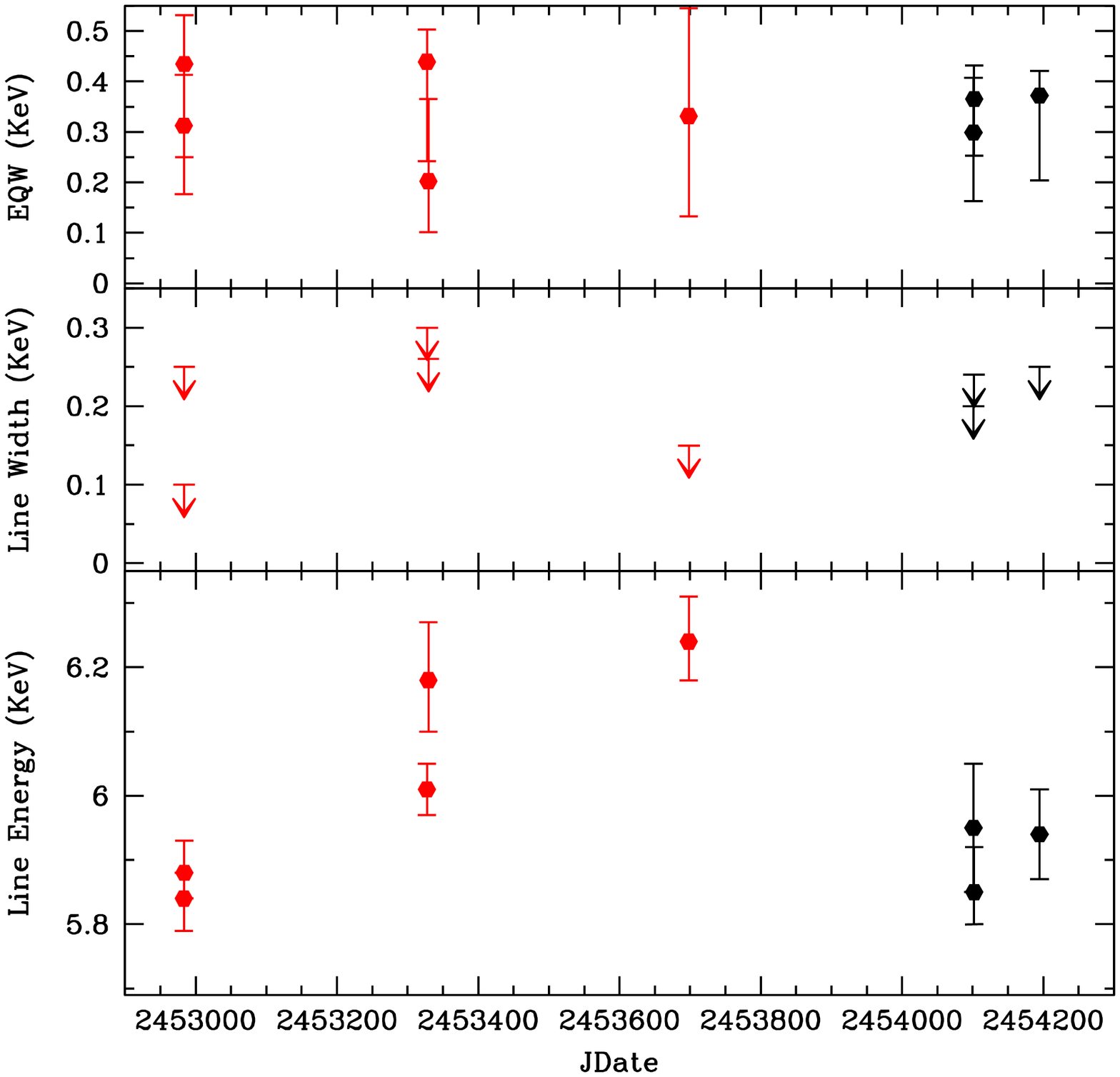}
\caption{\small Absorption line parameters (equivalent width, width and rest
frame energy of the absorption line) measured in different epochs spectra. Red
symbols represent \xmm\ observations (yrs. 2003, 2004 and 2005), black are
\chandra\ observations (yrs. 2006 and 2007). Error bars on the EW are at
1$\sigma$.}
\label{lc}
\end{figure}

In order to check the nature of the broadening of the absorption line, we analyzed the spectra
individually. Given that the \xmm\ data come from 8 observations taken over
three years, variability may be responsible for the broadening. Nine (5 \xmm\
and 4 \chandra) out of the 14 available observations have enough counts ($>$125)
to perform an individual spectral fit. The absorption feature is present at 2 to 3 $\sigma$
in all the spectra, apart from 1 \chandra\ observation where, although the
number of counts is enough for the fitting (187 net counts), the continuum flux
is faint and thus the absorption line is not detected. The intensity (with 1
$\sigma$ error), width and rest-frame energy centroid of the line measured
in the 8 spectra are plotted in Figure~\ref{lc} (red=\xmm, black=\chandra). In
contrast to the summed spectrum, the absorption features show a narrow profile,
with only upper limits on the width in all the spectra ($\sigma <$ 250 eV) and
the EW is consistent with being constant, implying a column density of gas of
the order of $\sim$5$\times$10$^{23}$cm$^{-2}$ (Bianchi et al. 2005).
 
The line energy centroid in the \xmm\ spectra changes in time over the range 5.9-6.3~keV, 
while in the \chandra\ spectra it is almost the same, within
the errors. This variation explains the broadening measured in the total \xmm\
spectrum, as due to the
superposition of narrow lines with different energy centroids. The variability 
of the energy centroid is of $\Delta$E=+500 eV in the
first 2 years and $\Delta$E=-400 eV in the last year. These shifts are comparable to
the 400 eV separation of the FeI and the FeXXVI K$\alpha$ absorption lines, so
a change in ionization could be in principle be responsible for the observed
shifts. We analyzed the
hardness ratio in all the spectra to check for column density variability to
be related with variability of the ionization state, but, given the large error
bars, we do not detect statistically relevant variability.

Instead, if this variation is
converted in a pure velocity variation, it implies a range in velocity of 0.02-0.07$c$
for neutral iron (Fe I), with a deceleration of 10,000 km/s/yr in the first two
years and a faster
acceleration in the last year of observations. If the lines
are due to completely ionized iron (Fe XXVI) absorption, the velocity is higher
(0.09-0.14$c$) and the implied acceleration is slightly lower.

Because of the degeneracy between velocity and ionization state, the same conclusions
on the absorber parameters (density, ionization state and velocity) are obtained
doing the single line analysis and the fitting with a photo-ionized model. 

\begin{table*}
\centering
\small
\caption{ Joint \xmm\ pn and \chandra\ ACIS X-ray spectral analysis results. 2 $\sigma$ errors are reported for 
all the quantities except for the EW of the lines for which we report 1 $\sigma$ errors. }
\label{Xray}
\begin{tabular}{ccc}
\hline\hline\\
\multicolumn{1}{c} {Fit Parameter}& \multicolumn{1}{c} {\chandra}& \multicolumn{1}{c} {\xmm}\\
\hline\\ 
Continuum & & \\
Mekal KT (KeV) & $<$ 0.13 & $<$ 0.13 \\
Mekal F$_{0.3-2}$ (erg s$^{-1} cm^{-2}$) & $<1.2\times10^{-15}$ &6.4$\times10^{-15}$ $_{-2.3}^{+5.2}$ \\
Photon Index $\Gamma$ & $1.95_{-0.06}^{+0.07}$ & $1.95_{-0.06}^{+0.07}$ \\
N$_{H}$ ($10^{22} cm^{-2}$) & $<$0.02 & $<$0.02 \\
\hline
Emission Line & & \\
Obs Energy (keV) & 6.44$_{-0.07}^{+0.07}$ & 6.60$_{-0.12}^{+0.15}$ \\
Line $\sigma$ (keV) & $<$0.57 & $<$0.20 \\
Flux (erg s$^{-1} cm^{-2}$) & 1.50$\times10^{-15}$ $_{-0.75}^{+1.14}$ & 1.35$\times10^{-15}$ $_{-0.70}^{+0.90}$ \\
EW (eV) & 570 $\pm$260 & 142$_{-86}^{+143}$ \\
\hline
Absorption Line & & \\
Obs Energy (keV) & 5.86$_{-0.14}^{+0.27}$ & 6.01$_{-0.14}^{+0.28}$ \\
Line $\sigma$ (keV) & $<$0.50 & 0.22$_{-0.11}^{+0.33}$ \\
Flux (erg s$^{-1} cm^{-2}$) & -5.68$\times10^{-16}$ $_{-3.36}^{+4.76}$ & -4.07$\times10^{-15}$ $_{-1.93}^{+2.23}$ \\
EW (eV) & -200 $_{-95}^{+98}$ & -441 $_{-148}^{+150}$ \\
\hline
Flux and Luminosity (\cgs)& & \\
Total F$_{0.5-10}$ & 4.49$\times10^{-14}$ $_{-0.20}^{+0.30}$ & 10.90$\times10^{-14}$ $_{-0.52}^{+1.20}$ \\
Total L$_{0.5-10}$ & 1.93$\times10^{43}$ & 4.85$\times10^{43}$ \\
Total F$_{0.5-2}$ & 1.89$\times10^{-14}$ $_{-0.10}^{+0.15}$ & 4.80$\times10^{-14}$ $_{-0.25}^{+0.34}$ \\
Total L$_{0.5-2}$ & 7.90$\times10^{42}$ & 2.18$\times10^{43}$ \\
Total F$_{2-10}$ & 2.60$\times10^{-14}$ $_{-0.20}^{+0.25}$ & 6.10$\times10^{-14}$ $_{-0.35}^{+0.44}$ \\
Total L$_{2-10}$ & 1.14$\times10^{43}$ & 2.66$\times10^{43}$ \\
\hline
\hline

\end{tabular} 
\end{table*} 

\section{Discussion}

We report in this paper on the properties of a peculiar and unique source at z=0.359, 
CID-42, in a survey of X-ray sources with ACS counterparts in COSMOS. 
CID-42 is a unique source in three specific ways. 

First, in the 2
deg$^2$ COSMOS area, there are $\sim$4 X-ray sources out of $\sim$550 at z$<$0.8
with optical morphologies indicative of ongoing major mergers, but only one of
them (CID-42) clearly shows two cores resolved in the HST image embedded in the
same galaxy but unresolved in X-rays. The surface brightness decomposition applied to the
ACS image strongly suggests the presence of a point like unresolved source (SE
source) and a more extended, but still compact, source (NW source).
The tidal tails to the NE and SW, in the ACS image, are evidence of merging
(Toomre \& Toomre 1972, Dubinski et al. 1999, Springel \& White 1999). The
morphology is consistent with either a early passage state, in a phase subsequent to
the first passage, or with a later stage of the merging after the coalescence of the two BHs (Mihos \&
Hernquist 1996, Cox et al. 2006, Hopkins et al. 2008).

Second, in 3 optical spectra obtained with two different instruments (VIMOS and
IMACS) a velocity offset larger than 1000 km/s is measured between the broad and
narrow component of H$\beta$ line. Large velocity offsets between two
different line systems can be observed if two BHs in a virialized
system are rapidly rotating very close to each other (sub-parsec separation), as
suggested for the double line system of Boroson \& Lauer (2009).  
However, assuming that the broad and narrow lines are coming from the SE and NW
sources respectively, virialization is not appropriate for the kiloparsec scale 
separation of the two sources in CID-42. 

A possibility to justify the
large velocity offset is a recoil effect. Large recoiling velocity have been
predicted by theoretical models and simulations for both GW wave kicks (Campanelli et
al. 2007a, b) and slingshot effect kicks in triple BH systems (Hoffman \& Loeb 2007). 
The unresolved morphology of the SE source allows us to assume that it plays the
role of the unobscured SMBH (a Type 1 AGN) and produces the broad H$\beta$ line. 
Therefore the SE source is recoiling with respect to the system with a velocity
of $>$1000 km/s. 

As the BH recoils from the center of a galaxy or from another BH, the closest, most tightly
bound, regions (disk and BLRs) are carried with it, and the more distant 
regions (NLRs) are left behind (Loeb 2007). 
Given the distance between the recoiling BH and the center of the whole system 
(measured on the plane of the sky) and one component of the BH velocity (the
radial velocity) with respect to the galaxy, it is possible to compute
a lower limit on the time since the BH has been kicked of
$\sim$4$\times$10$^6$ yr. 

The detection of a recoiling BH is related to its lifetime and it is 
extremely rare because it never gets large separation before fading. 
The lifetime of the active phase of the BH depends on the amount of accreting
material it carries after the kick. In principle, higher velocities result in
a smaller ejected disc. In CID-42, the BH is still emitting enough radiation
to ionize the BLR, so it had to carry enough
material in order to be still active. 

It is possible to compute the BH mass of SE source using the relation of 
Vestergaard \& Peterson (2006)
between the line width, the BH mass and the continuum luminosity, by using the
H$\beta$ line width (3500$\pm$233 km/s). As shown in Figure
\ref{sed}, the total continuum luminosity is mostly dominated by the galaxy
light. Taking advantage of the surface brightness decomposition results, we can
extrapolate the 5100 \AA\ nuclear luminosity of
$\sim$4$\times$10$^{43}$~\lum by normalizing a standard quasar
template (red line in Fig.~\ref{sed}, Elvis et al. 1994) to the SE
source luminosity in the ACS filter (square in Fig.~\ref{sed}). Under this
assumption, the BH mass of the SE source is $\sim$6.5$\times$10$^7$
M$_{\odot}$.

Given the total mass of the galaxy and the estimated BH mass, the ratio
M$_{BH}$/M$_{\star}$ is a factor 2 higher than the H{\"a}ring \& Rix (2004)
M$_{BH}$-M$_{\star}$ relation for local spheroids. Although the errors on the 
BH and galaxy mass are quite large, this is consistent with
the fact that CID-42 is a peculiar source or that the local relation
deviates at high z (see for example Treu et al. 2007 and Merloni et al. 2010). If we consider
that the galaxy mass is an upper limit to the bulge mass, the offset from the
relation could be even bigger. Using the bolometric luminosity and BH mass of
the SE source, we find that the Eddington ratio is L/L$_{Edd}\sim$0.04, 
consistent either with the nuclear activity of this source having just been
triggered or already declining. Using the correlation between the
Eddington ratio and the bolometric correction in the 2-10 keV energy range
(Vasudevan \& Fabian 2009, Young et al. 2010), it is possible to estimate the
2-10 keV luminosity emitted by the SE source to be in the range
0.6-1$\times$10$^{43}$ \lum, consistent with the 2-10 keV luminosity measured for
CID-42 (see Table~\ref{Xray}).

The third feature which makes CID-42 unique is the detection, in the X-ray spectra, 
of an extraordinarily strong, {\it
redshifted} absorption iron line and an emission iron line, forming an {\it
inverted} P-Cygni profile. Out of the 100 X-ray sources with 
more than 1000 counts in the stacked X-ray spectrum only CID-42 shows an iron line 
in absorption (Mainieri et
al. 2007, Lanzuisi et al. 2010 in prep.). The absorption line is intrinsically
narrow. Its energy centroid changes by 500 eV over the 4 years of
observations. The broadening of the absorption line in the total \xmm\ spectrum
is the result of the stacking of narrow lines redshifted by different velocities
in the different observations.
Such {\it redshifted} absorption lines are rare. Most known X-ray absorbers in
AGNs are
blueshifted (see Cappi 2006 for a review) and 85\% of them are at low redshift
(z$<$0.2). As blueshifted absorbers necessarily imply out-flowing winds,
redshifted absorbers would seem to require high velocity inflows, which must
therefore be situated very close to the central black hole. The few
convincing cases of redshifted lines are all narrow absorption lines detected in
Seyfert 1 sources (Mrk 335, Longinotti et al. 2007; PG 1211+143, Reeves et
al. 2005; Mrk 509, Dadina et al. 2005), and have suggested explanation as gas falling
with relativistic velocity directly into the BH at a few tens of
gravitational radii.

The cluster
of rare features seen in this source, more likely associated with a brief moment
of the merging event, is a strong argument against the chance superposition hypothesis. 
Any model for the system should
preferably link these unique features. We describe two possible models
below.

\subsection{Gravitational Wave Recoiling Black Hole}

The spatial offset of the SE point-like source with respect to the NW source and
the kinematic offset of the broad H$\beta$ line with respect to the narrow line
system can be explained by a SMBH ejected by gravitational
radiation produced in the black hole merger.

According to theoretical models, during the final coalescence of two BHs in 
a merger event, the gravitational waves produced can impart a large kick 
to the resultant BH, which recoils in the opposite direction with
respect to the center of the galaxy (Peres 1962). Although some approximate
methods indicated that the kick velocities are most likely small
($<$few$\times$100 km/s, Blanchet et al. 2005, Damour \& Gopakumar 2006), the
results from full numerical relativity simulations show that the recoil speed
can be up to 4000 km/s (Campanelli et al. 2007a, b). 

It is difficult to estimate how often significant recoil kicks occur, because
the recoil kick distribution depends on the spin and mass ratio distributions of
the progenitor BH binaries, which are not well known. Several authors
(Schnittman \& Buonanno 2007, Campanelli et al. 2007a, b and Baker et al. 2008)
estimated, using results from full numerical relativity simulations, that
12-36\% of recoiling BHs will have velocities $>$ 500 km/s and 3-13\% $ >$ 1000
km/s, assuming a spin of 0.9 with random orientations and using different
assumptions on how the kick velocity scales with BH progenitor mass ratio
(0.1$<q<$1). These fractions might be smaller if BH spins are on average lower
or if gas torques are efficient at aligning the spin axes in a way that is
unfavorable for large kicks (Bogdanovic et al. 2007, Dotti et al. 2009). To
constrain the actual recoil kick distribution, better observational constraints
on merging BHs or observations of individual recoiling BHs are required. A
candidate recoiling SMBH with a large kick velocity of 2650 km/s has been
spectroscopically discovered by Komossa et al. (2008), although alternative
models have been proposed to explain the two systems of lines in the optical
spectrum. A second possible candidate has been also proposed by Shields et
al. (2009). 

The high velocity offset (v$>$1000 km/s) measured in the optical spectra of
CID-42 is consistent with the numbers obtained in simulations, although the probability
of finding such high velocity recoiling BH is low ($<$10\%).

The surface brightness decomposition of the CID-42 ACS image is consistent with
the ejected BH hypothesis too: the SE point-like source, the ejected BH, is
spatially separated from the compact, but not point-like, NW bright source,
which can easily represent the
galaxy core from which the BH has been ejected.

The fact that the broad lines are redshifted with respect to the
stellar absorption lines is consistent with the recoiling BH picture, because they are
following the BH, as explained at the beginning of Section~5. The narrow lines in 
the spectra have the same redshift of the
stellar absorption lines but, given that the NLRs, residing in the NW source, are
not powered anymore by the BH and will fade long before the ejected BH reaches
kpc distances, another way to produce the narrow lines is
needed. The line flux ratios indicate that the emission is produced by nuclear
activity and not by star formation (Comerford et al. 2009). 
The narrow lines could be produced by the ejected BH ionizing the local galaxy ISM on its way
out from the center. In this case, the narrow lines should have the same
redshift of the galaxy, as observed, implying that the BH is still inside a
region of the galaxy disk with high ISM density or is passing by
molecular clouds. In this case, the velocity offset measured
in the optical spectrum, not corrected for instrumental shift (see Section 4.1),
is 1360$\pm$320 km/s, which represents the radial component of the total
velocity.

Lousto et al. (2009) showed in a statistical studies of black-hole binaries
during the inspiral and merger, that when the kick velocity
is $\gtrsim$1000 km/s, the direction of the kick is most likely perpendicular to
the BH orbital plane. They found a probability of 6.3\% for large recoil
velocities ($>$1000 km/s) along the line of sight. 
If the accretion disk is aligned with the nearly face-on galaxy disk as a whole,
then a large kick closely aligned with the line of sight is expected, and it is
consistent with what observed in CID-42, if the source is a recoiling BH.

The time since the BH has been kicked ($\sim$4$\times$10$^6$ yr), as computed at
the beginning of Section~5, is in agreement with the prediction of the effective time for
the persistence of off-center quasar activity (10$^{6.2-7}$ yr, Blecha \& Loeb
2008). This number is a lower limit but it should be a good estimate, assuming that the
total kick velocity is not much higher than the radial velocity, given that this
is already in the extreme tail of the kick velocity distribution predicted by
models and simulations. According to Blecha \& Loeb (2008) model,
for a kick velocity in the range 1100 - 1500 km/s and a BH mass of 10$^7$-10$^8$
M$_{\odot}$, the mass of the BH disc, in order to be still emitting, is $\sim$2-3\% of the BH mass.

Models predict that the effect on galaxy morphology of BH recoil is to produce 
significant asymmetry (Kornreich \& Lovelace 2008). 
This is seen in the morphology
of CID-42 where the galaxy core (the NW source) is slightly misplaced with
respect to the geometric center of the galaxy disc.

The variable Fe-K absorption line can be alternatively explained,
as already explained at the beginning of this section, by material
inflowing into the BH, as it has been proposed in a few other sources (Cappi 2006, see Section~5). 
In order to have such a
high velocity, the inflow of material would have to be very close to the nucleus (few
tens of Schwarzschild radii). One means of achieving high inflow velocity is
with a photon bubble instability (Arons 1992, Gammie 1998, Begelman 2006) which
produces a propagating pattern of low and high density regions through which the
radiation has to go. The photon bubble instability is currently used to explain 
the {\it inverted} P-Cygni profiles detected in the spectra of luminous blue 
variable (LBV) stars (e.g., van Marle et al. 2008). Although this would be
a physical explanation, it requires super-Eddington accretion, that is not
observed in CID-42.

Although the high velocity inflow possibility is very interesting, without a detailed
modeling it remains mere hypotheses.

\subsection{A Type 1 AGN recoiling from a Type 2 AGN in a slingshot}

In the following we list the evidences in favor of a picture in which 
CID-42 contains 2 AGNs, an unobscured Type 1 AGN in the SE source and an obscured
Type 2 AGN in the NW one.

\begin{enumerate}
\item The surface brightness decomposition of the ACS image finds 
a compact, but not point-like, bright source in the NW nucleus.
\item The shift ($\Delta v \sim$1180 $\pm$360 km/s) between the broad and narrow
components of the H$\beta$ line in the optical spectrum also suggests the presence of
two active nuclei, one responsible for the broad line emission (i.e. a Type 1
AGN), the second responsible for the narrow emission lines only (i.e. a Type 2
AGN). A possible explanation for the presence of only one narrow emission line
system could be that the narrow line regions (NLRs) of the two sources are mixed up, and,
for this reason, are at the same redshift measured for the galaxy from the
absorption lines. However, we discard this hypothesis as NLRs, at the
luminosity of CID-42, are expected to extend 0.1-1 kpc from the nucleus
(Schmitt et al. 2003), much less than the projected separation measured
between CID-42 nuclei.
\item The X-ray flux variability, and the presence of a strong iron emission
line (EW$\sim$500 eV) when the continuum is weak also suggest a combination of a
variable Type 1 AGN and a constant obscured Type 2 AGN. Usually, strong iron
emission lines (EW$\sim$1keV) are associated with heavily obscured (Compton
Thick) sources and are less prominent in unobscured ones (EW$\sim$150 eV; e.g.,
see Figure 9 of Guainazzi et al. 2005a). As already noted by Guainazzi et
al. (2005b), all the AGN pairs observed so far are found to be heavily obscured
or Compton thick in the X-rays, in both components. The presence of at least one
obscured AGN in our system is consistent with this picture but the possibility
that we have a Type 2 plus a Type 1 AGN embedded in the same galaxy is new, although 
we believe CID-42 is not a binary.
\end{enumerate}

Assuming that the [OIII] emission lines are produced by the Type 2 AGN only, it
is possible to roughly estimate the contribution of the two sources to the total
X-ray luminosity. The Type 2 luminosity is estimated from the relation between
the L$_{2-10 keV}$-L$_{[OIII]}$ measured in the COSMOS field (Silverman et
al. 2009) and the Type 1 AGN luminosity is then found by subtraction. Given the
[OIII] luminosity ($\sim$3$\times$10$^{41}$ \lum) and the uncertainties on this
relation, the luminosity of the NW Type 2 nucleus would be $\sim$9$\times$10$^{42}$
\lum\ in the 2-10 keV band. The derived Type 1 AGN luminosity (2-10 keV) is in 
the range 0.3-1.8$\times$10$^{43}$ \lum, due to the strong measured variability
(Fig. \ref{sed}, right). The last number is consistent with the 2-10 keV
luminosity obtained in the previous section with the Eddington ratio.

As explained at the beginning of Section 5, the large velocity shift between 
the two nuclei implied by the optical spectra would require a large mass of the 
system assuming it is virialized (i.e. the two nuclei are gravitationally bound in a circular orbit).  
Comerford et al. (2009) detected two sets of [OIII] and
H$\alpha$ lines in a Keck/DEIMOS spectrum\footnote{Although the spectral
resolution of the Comerford et al. (2009) spectrum is higher than that of the IMACS and
VLT spectra one, the signal-to-noise ratio, estimated from their Figure 3, is
lower.}, from which a $\Delta$v=150 km/s is measured by detecting two sets of
emission line. This lower velocity shift allows a more ``reasonable'' mass for a
virialized system, however, virialization, which is a fair assumption for close
pairs, is probably not appropriate for the kiloparsec
scale separation of the two CID-42 sources. Most likely, the
velocity measured by Comerford et al. (2009) is due to the rotation of the gas
in the galaxy as a whole, as the system is in a violent state of gravitational
interaction. 

The large velocity offset between the two source is more likely due to a nearly 
radial orbit, with the Type 1 SE AGN is recoiling with respect to the Type 2 AGN. 
A velocity difference of this size may result from 
a slingshot ejection in a triple BH encounter (Saslaw et al. 1974, Hoffman \& Loeb 2006, 
2007). If a BH binary stalls for 
a time of $\sim$ few Gyrs before it coalesces, due to the depletion of stars in
the loss cone in a gas poor merger (Merritt \& Milosavljevic 2005), 
its host galaxy could merge with another galaxy and a
third BH may spiral in and undergo a strong 3-body interaction with the binary. 
The just arrived system will merge with the original galaxy. The star formation
rate ($\sim$100 ~M$_{\odot}$/yr) measured from the SED fitting, supports the 
scenario of a merger in a early phase.
The merger will drive mass and gas to the binary, refilling its loss cone. 
The configuration will become gravitationally unstable and 
this event will end with the further reduction of the separation or the 
coalescence of the binary and the ejection of the
lightest body at a speed comparable to the binary's orbital speed (Heggie 1975).

The large projected distance (2.4 kpc) between the two sources and the high 
velocity measured in in CID-42 are possible for this kind of ejection 
(Hoffman \& Loeb 2006, 2007), although the probability to have a certain
velocity depends from the initial condition of the system and a proper 
modeling is still undergoing.

\subsubsection{Backlit wind model}

\begin{figure}[t]
\includegraphics[width=0.5\textwidth]{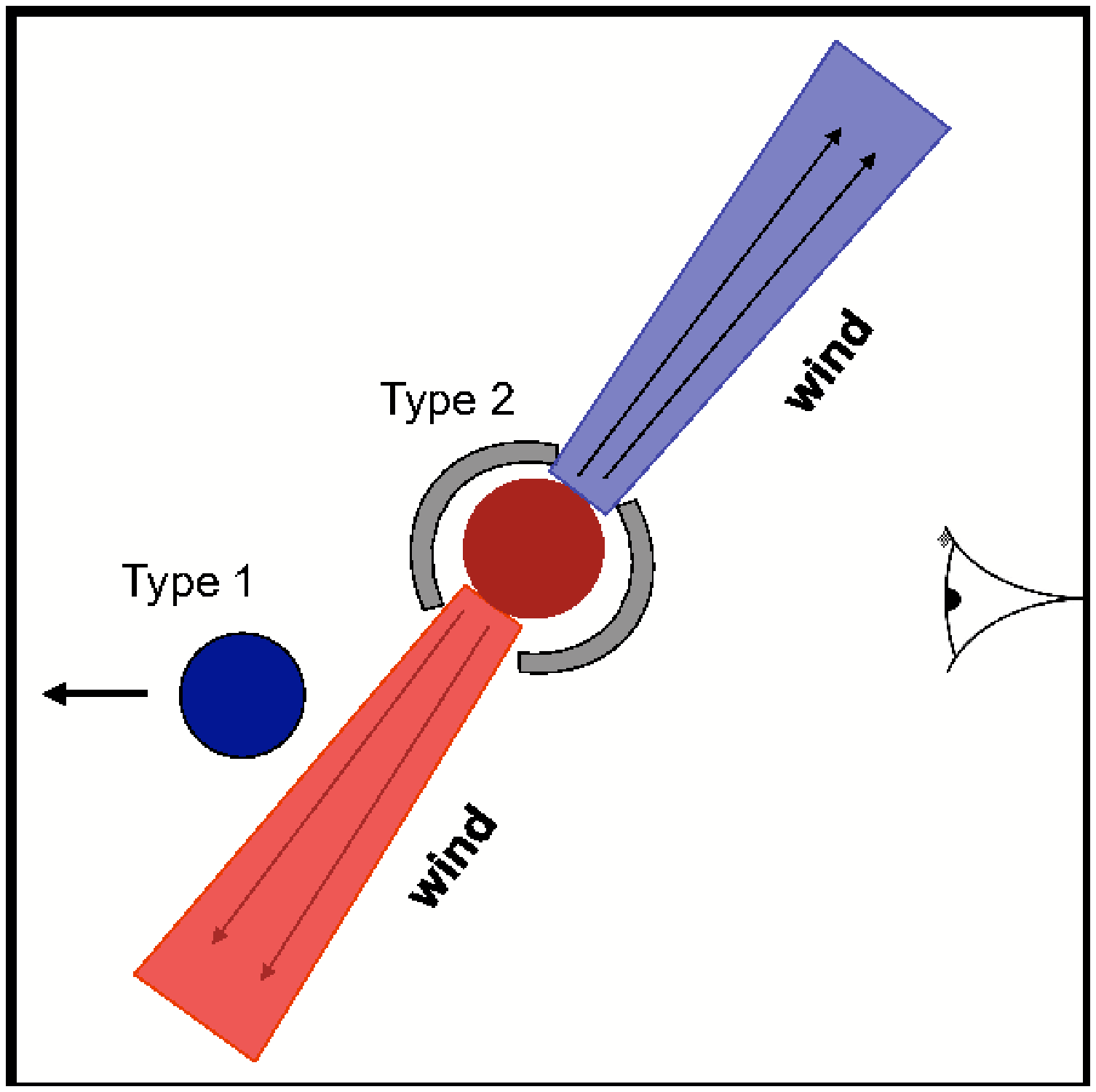}
\caption{\small Backlit Wind model sketch. The proposed geometry for the system
is presented. In red the Type 2 AGN (NW nucleus) with the absorbing material
around (gray), in blue the Type 1 (SE nucleus). The component of the BAL-like
wind, which block the line of sight towards the Type 1 AGN, is shown as a red
cone. The blueshifted wind (blue cone) is unobservable. The line of sight is
presented with a symbolic eye. }
\label{model}
\end{figure}

On the basis of the above analysis, the SE nucleus is the unobscured Type 1
recoiling AGN,
while the NW is an obscured Type 2 AGN, resulting from the coalescence of a
binary. The two nuclei lie close to the plane
of the sky possibly moving on a radial orbit in a direction close to the line of
sight, though a small angle has to be considered given the 2.4 kpc separation. 
Relative to the mean velocity of the system, the SE nucleus is moving
away from the observer and the NW one towards the observer. The suggested
geometry of the system is shown in Figure~\ref{model}.

The SE Type 1 nucleus then has: a strong unobscured optical continuum and BLR,
leading to the point-like HST/ACS image, the broad H$\beta$ emission line and a
strong, variable, X-ray continuum emission. The obscured NW Type 2 nucleus has: a
weak optical continuum, leading to a compact but extended HST/ACS source
dominated by the galaxy light, only narrow emission lines of H$\beta$ and
[OIII], and a weak X-ray continuum, with a strong fluorescent iron emission line
(EW$\geq$ 570 eV).

The proximity of the two sources allows a special interpretation of the Fe-K
absorption line as a ``Backlit Wind''. In this scenario, a fast, highly ionized
BAL-like wind is emitted by the NW Type 2 nucleus (located in front). In almost
all circumstances this wind would be undetectable. However, in the case of
CID-42, the {\em receding} wind crosses our line of sight to the SE Type 1
nucleus, leading to absorption of the Type 1 continuum in redshifted iron. The
blueshifted emission of the approaching wind remains invisible as there is no
background source to illuminate it. The column density of the flow derived from
the EW of the absorption line is consistent with what is seen in BAL wind
($>$10$^{23}$ cm$^{-2}$; Green et al. 2001, Gallagher et al. 2002, Chartas et
al. 2009). The fact
that we measure no X-ray photoelectric absorption at low energies implies that the
wind is highly ionized, something observed also in other BAL (e.g., Telfer et
al. 1998, Hamann 1998, Chartas et al. 2009) and X-ray absorbers (e.g., Vignali et al. 2000,
Reeves et al. 2009).

The variations of the iron absorption line energy centroid on years timescale can be
explained either by velocity variations ($\sim$10,000 km/s per year) in the
flow, or by a change in the ionization state, and so in the density (Proga et al. 2000)
or, possibly, to a precession of the flow, so that a change in inclination can
produce a difference in velocity, or by a combination of all these effects.
BALs, where we look directly along the flow, are known to vary on year
timescales (e.g., Lundgren et al. 2007, Gibson et al. 2008), so variability
should be more likely in a section viewed across the flow. The
variability implies that the BAL-like flow must be highly structured, even 2.5
kpc away from the nucleus. This is not unusual as it has been shown by \chandra\
detection of radio jets still relativistic at Mpc distance from their origin 
(e.g., PKS 0637-752, Schwartz et al. 2000).

As we have only one example of {\it redshifted} iron absorber and only one
recoiling BH, this scenario is only plausible if the BAL-like wind has a large
opening angle, making an interception of the rear recoiling Type 1 nucleus likely.

Winds at a few 1000 km/s are normal in Type 1 AGNs (Reynolds 1997, Piconcelli et
al. 2005) and much faster (0.1-0.2$c$) BAL winds, confined to modest solid
angles, are not
unusual (Weymann et al. 1991, Hamann et al. 1993, Ogle
et al. 2000). Fast winds at kpc-scale distances from Type 1 AGN have been found 
as well (Arav et al. 2001) and have been proposed to be
universal in AGNs (Elvis 2000). However, there has been no possibility to test
this idea in Type 2 AGNs. CID-42 provides this opportunity and gives us the
first detection of a wind in a Type 2 AGN.

\section{Conclusions}

We have reported the multiwavelength properties of CID-42, 
which presents three unusual features: two close sources, embedded in the same galaxy, 
resolved in the optical ACS image, but unresolved in the X-ray one,
a high
velocity offset, measured between the broad and narrow H$\beta$ lines in 3
different optical spectra and a {\it redshifted} absorption iron line in the X-ray spectrum.

Thanks to the rich database of the COSMOS survey, the analysis of the properties
of CID-42 has been performed at all wavelengths.

The overall analysis suggests two possible explanations for the source: 
a GW recoiling BH, caught $\sim$few Myr after the kick
from the center of the galaxy or a Type
1/Type 2 AGN system where the Type 1 is recoiling due to a slingshot effect.

Analysis of the ACS image shows the presence of a SE point-like source,
playing the role of the Type 1 AGN (or SMBH with disc and BLRs) in both scenarios, confirmed
by the X-ray flux variability across the time and by the offset in velocity
measured between the narrow and the broad components of the H$\beta$ line. The
NW source, being less compact than the SE one, could be either the
naked core of the galaxy from which a BH has been kicked out or a Type 2 AGN.

The presence of a {\it redshifted} absorption iron line changing its energy centroid
in the different observations, allows to explain the geometry of the system 
by inflow of material into the BH in the GW recoil BH case or with
the ``Backlit Wind'' model in the second scenario. Monitoring observations in the X-ray would be
suitable to study the moving X-ray absorption feature and its variability.

The ``Backlit Wind'' implies that fast BAL-like winds are present in Type 2
AGNs, an otherwise untestable hypothesis. The variability of the iron absorption
is, in this special case, a new tool to study BAL flows. This model predicts
corresponding high ionization UV absorption lines (e.g. OVI), making CID-42 a suitable
target for HST/COS observations. 
\chandra\ high-resolution observations with HRC could resolve the two sources
to see if both are X-ray emitting or not, while optical and IR
spectroscopy with higher spatial resolution, which currently is only possible
with HST, could confirm the still uncertain velocity shift measured in
H$\beta$ in other lines.

If future observations allow us to confirm that the GW recoiling BH is the
best explanation of the system, it will be the first time in which this
phenomenon is observed via both spatially offset from the galactic core and a
velocity offset in the optical spectrum.

Although models for this system have been sketched qualitatively in this paper, a
numerical analysis will be the subject of a forthcoming paper. 

Upcoming Keck Integral Field Unit (IFU) observations with OSIRIS will 
help us to disentangle major competing models mapping the inner region of
CID-42. IFU observations on larger scales could map the velocity
of gas and stars in the galaxy and in the tails, in order to constrain 
the theoretical models on the geometry of the merger and
the inclination of the galaxies in the first encounter (D'Onghia et al. in
prep.).

Calibrating the number of recoiling BH, produced either in close triple
encounters and by gravitational waves ejection could be of great importance for
the proposed Laser Interferometer Space Antenna (LISA) mission (see discussion in Loeb 2009).
 
CID-42 is a unique source with a cluster of rare features. Whichever is the
best explanation, in CID-42 we are witnessing a runaway BH.

\section{Acknowledgements}

FC thanks E. Costantini, T.J. Cox, M. Dotti, L. Hernquist and A. Sesana for useful discussion. 
The authors thanks Chien Y. Peng for useful discussions about GALFIT.
The authors thank the anonymous referee, whose critical analysis helped to improve this 
paper, making it more interesting. 
This work was supported in part by NASA {\em Chandra} grant number
GO7-8136A (FC, ME, AF), NASA contract NAS8-39073 (Chandra X-ray Center). 
KJ acknowledges support from the Emmy Noether Programme of the German
Science Foundation (DFG) through grant number JA~1114/3-1.
In Italy this work is supported by ASI/INAF contracts I/023/05/0,
I/024/05/0 and I/088/06, by PRIN/MUR grant 2006-02-5203. In Germany
this project is supported by the Bundesministerium f\"{u}r Bildung und
Forschung/Deutsches Zentrum f\"{u}r Luft und Raumfahrt and the Max
Planck Society.

{\it Facilities:} \facility{\chandra\ (ACIS)}, \facility{HST (ACS)}, \facility{\xmm}.

\end{document}